\documentclass[preprint,aps,prd,floatfix,nofootinbib,11pt]{revtex4-2}
\pdfoutput=1
\usepackage{graphicx}
\usepackage{amsfonts}
\usepackage{amssymb}
\usepackage{xcolor}
\usepackage{natbib}
\usepackage{braket}

\usepackage{hyperref}

\usepackage{amsmath}
\usepackage{rotating}
\usepackage{amssymb,amsthm}
\usepackage{soul}

\usepackage{xcolor}

\usepackage{latexsym}
\usepackage{graphics}

\usepackage{amsfonts}
\usepackage{amsmath}
\usepackage{rotating}
\usepackage{amssymb}

\usepackage{amsmath}
\usepackage{rotating}
\usepackage{amssymb}
\usepackage{soul}

\usepackage{xcolor}
\usepackage{xcolor}
\def\beq{\begin{eqnarray}}  
	\def\eeq{\end{eqnarray}}

\begin{document}
	
	\title{Scalar Field Static Spherically Symmetric Solutions in Teleparallel $F(T)$ Gravity}

	\author{A. Landry}
	\email{a.landry@dal.ca}
	\affiliation{Department of Mathematics and Statistics, Dalhousie University, Halifax, Nova Scotia, Canada, B3H 3J5}

%add authors here

\begin{abstract}
We investigate in this paper the static radial coordinate-dependent spherically symmetric spacetime in teleparallel $F(T)$ gravity for a scalar field source. We begin by setting the static field equations (FEs) to be solved and solve the conservation laws for scalar field potential solutions. We simplify the FEs and then find a general formula for computing the new teleparallel $F(T)$ solutions applicable for any scalar field potential $V(T)$ and coframe ansatz. We compute new non-trivial teleparallel $F(T)$ solutions by using a power-law coframe ansatz for each scalar potential case arising from the conservation laws. We apply this formula to find new exact teleparallel $F(T)$ solutions for several cases of coframe ansatz parameter. The new $F(T)$ solution classes will be relevant for {studying the models close to Born--Infeld and/or scalarized Black Hole (BH) solutions inside the} dark energy (DE) described by a fundamental scalar field such as quintessence, phantom energy or quintom system, to name only those types.
\end{abstract}

\maketitle

\newpage

\section{Introduction}

{The} %MDPI: We removed font colors in the whole paper, please check.
Teleparallel $F(T)$ gravity {is a promising frame-based and alternative theory to general relativity (GR) fundamentally defined in terms of the {coframe} %MDPI: 1. Please check if the bold format for variables is necessary. If not, please remove. Please check all bold variables and make sure that all the symbols in the paper are of the same format. AL: It's OK for bold variable here.
	${\bf h}^a$ and the spin-connection $\omega^a_{~bc}$~\cite{Lucas_Obukhov_Pereira2009,Aldrovandi_Pereira2013,Bahamonde:2021gfp,Krssak:2018ywd,MCH,Coley:2019zld,Krssak_Pereira2015}. These two last quantities define the torsion tensor $T^a_{~bc}$ and torsion scalar $T$. {We cannot forget that GR is defined by the metric $g_{\mu\nu}$ and the spacetime curvatures $R^a_{~b\mu\nu}$, $R_{\mu\nu}$ and $R$.} The teleparallel gravity features are the new possible spacetime symmetries under non-trivial linear isotropy groups and the Lorentz-invariant geometries~\cite{chinea1988symmetries,estabrook1996moving, papadopoulos2012locally}. Under this consideration, we can determine the symmetries for any independent coframe/spin-connection pairs, and then spacetime curvature and torsion are defined as geometric objects~\cite{MCH,Coley:2019zld,olver1995equivalence}. From~here, any geometry described by a such pair whose curvature and non-metricity are both zero ($R^a_{~b\mu\nu}=0$ and $Q_{a\mu\nu}=0$ conditions) is a teleparallel gauge-invariant geometry (valid for any $g_{ab}$). The fundamental pairs} must satisfy two Lie derivative-based relations and we use the Cartan--Karlhede algorithm to solve these two fundamental equations for any teleparallel geometry. For a pure teleparallel $F(T)$ gravity spin-connection solution, we also solve the null Riemann curvature condition leading to a Lorentz transformation-based definition of the spin-connection $\omega^a_{~b\mu}$.

{The direct equivalent to GR in teleparallel gravity is} the teleparallel equivalent to GR (TEGR) {and then it generalizes to teleparallel $F(T)$-type gravity with a function $F(T)$ of the torsion scalar $T$ \cite{Aldrovandi_Pereira2013,Ferraro:2006jd,Ferraro:2008ey,Linder:2010py}. This generalization} is locally invariant under the covariant Lorentz definition~\cite{Krssak_Pereira2015}. {Beyond the teleparallel $F(T)$ gravity, all the previous considerations have been adapted for the new general relativity (NGR) (refs.~\cite{kayashi,beltranngr,bahamondengr} and references therein), the symmetric teleparallel $F(Q)$-type gravity (refs.~\cite{heisenberg1,heisenberg2,faithman1,hohmannfq} and references therein) and some extended theories like $F(T,Q)$-type, $F(R,Q)$-type, $F(R,T)$-type, and several other ones (refs.~\cite{jimeneztrinity,nakayama,ftqgravity,frqspecial,frtspecial,frttheory,myrzakulov1,myrzakulov2,myrzakulov3,myrzakulov4,myrzakulov5} and references therein). The most relevant is by using the teleparallel $F(T)$ gravity framework for the current development.}

{
	There are a large number of research papers on spherically symmetric spacetimes and solutions in teleparallel $F(T)$ gravity using a large number of approaches, energy-momentum sources and made for various purposes~\cite{golov1,golov2,golov3,debenedictis,SSpaper,TdSpaper,nonvacSSpaper,nonvacKSpaper,scalarfieldKS,roberthudsonSSpaper,coleylandrygholami,scalarfieldTRW,baha1,bahagolov1,awad1,baha6,nashed5,pfeifer2,elhanafy1,benedictis3,baha10,baha4,ruggiero2,sahoo1,sahoo2,calza}. However, the resolution of field equations (FEs) in the orthonormal gauge is performed for coframe/spin-connection pair determinations and to avoid the extra degrees of freedom (DoF) problem associated to the FEs on proper frames ({refs.~\cite{SSpaper,nonvacSSpaper,nonvacKSpaper,roberthudsonSSpaper}} for detailed discussions). However, the~symmetric FEs and the $F(T)$ solutions are similar for any gauges (orthonormal or not) confirming the gauge invariance of teleparallel $F(T)$ gravity FEs. By setting the orthonormal gauge, we solve the non-trivial antisymmetric and symmetric FE parts covering all DoFs for a non-trivial and orthonormal coframe/spin-connection pair (see refs.~\cite{SSpaper,nonvacSSpaper,nonvacKSpaper,roberthudsonSSpaper,scalarfieldKS} for detailed discussion). In~the literature, a large number of papers use tetrads $e^a_{~\mu}$ instead of coframes $h^a_{~\mu}$ for the orthonormal gauge by default definition~\cite{golov1,golov2,golov3,debenedictis,baha1,bahagolov1,awad1,baha6,nashed5,pfeifer2,elhanafy1,benedictis3,baha10,baha4,calza}. This very usual approach prevents the use of the non-orthonormal frame and constitutes its main weakness. To avoid this last limitation, to~respect the gauge invariance and to work on any type of frame not necessarily orthonormal, the coframe $h^a_{~\mu}$ is used here and has been used in refs.~\cite{SSpaper,nonvacSSpaper,nonvacKSpaper,roberthudsonSSpaper,scalarfieldKS,scalarfieldTRW} for solving the static spherically symmetric (radial coordinate dependent) teleparallel $F(T)$ gravity FEs in the case of linear and quadratic perfect fluid sources \cite{nonvacSSpaper}. There are also additional developments on cosmological and time-dependent spherically symmetric teleparallel $F(T)$ and $F(T,B)$ solutions leading to dark energy (DE) source models in cosmic background (refs. \cite{nonvacKSpaper,roberthudsonSSpaper,scalarfieldKS,coleylandrygholami,scalarfieldTRW,FTBcosmogholamilandry} for details). For the last reason, we will consider the static spherically symmetric teleparallel $F(T)$ solutions with a scalar field source to complete the most recent studies on this topic. Beyond the spacetime structures and possible teleparallel solutions, there are further physical justifications and a significant number of possible applications.}

The most important classes of applications to the static radial coordinate-dependent scalar field source teleparallel $F(T)$ solutions are the astrophysical problems (refs. \cite{awad1,bahagolov1,baha6,nashed5,pfeifer2,elhanafy1,benedictis3,baha10,baha4,ruggiero2,sahoo1,sahoo2} for good examples). This especially concerns the black hole (BH), Neutron Star (NS), White Dwarf (WD) and any system containing the previous types of astrophysical objects or any radial-centered systems. The last list is so far not exhaustive. The recent literature has shown that any astrophysical system is evolving in a DE-dominated universe \cite{steinhardt1,steinhardt2,steinhardt3,carroll1,quintessencecmbpeak,quintessenceholo,quintchakra2024,cosmofate,rollingscalarfield,steinhardt2024,wolf1,wolf2,wolf3,quintessencephantom,strongnegative,farnes,baumframpton,caldwell1,phantomdivide,phantomteleparallel1,ripphantomteleparallel2,phantomteleparallel3}. We have taken into account this fact in a recent work on teleparallel $F(T)$ gravity by considering a DE perfect fluids dominating the spacetime in ref. \cite{nonvacSSpaper}. Several recent papers in teleparallel cosmology have been developed in the scope of DE-dominating models \cite{leonpalia2,Kofinas,paliathanasis2022f,attractor,FTBcosmogholamilandry,roberthudsonSSpaper,nonvacKSpaper,scalarfieldKS,scalarfieldTRW}. We need to take into account DE perfect fluids and/or scalar fields for any astrophysical process in teleparallel gravity in the future. This relevant approach constitutes a more realistic manner to study the astrophysical processes in the universe. The primary aim of this paper is to find new classes of teleparallel $F(T)$ gravity solutions in the scope of treating astrophysical systems evolving in a DE fundamental scalar field. This paper naturally follows the work achieved and directly completes the analysis in ref. \cite{nonvacSSpaper}. But before going to the new possible teleparallel $F(T)$ solutions, we need to further clarify the types of~DE.

The DE and traditionally the scalar field quintessence behavior are usually studied by using the perfect fluid equivalent equation of state (EoS) $P_{\phi}=\alpha_Q\,\rho_{\phi}$ where $\alpha_Q$ is the DE index (or quintessence index in some references). According to this index, the possible DE forms can be summarized as follows:
\begin{enumerate}
	\item \textbf{{Quintessence} %MDPI: Please confirm if the bold is unnecessary and can be removed. The following highlights are the same. Please check all bold in the list format in the whole paper. AL: Bold is necessary because I make reference to these concept in the development of this paper.
	} {${\bf -1<}\alpha_Q {\bf <-\frac{1}{3}}$}%MDPI: Please also check if bold for numbers are necessary. or if variables should also be bold here? OK: Bold necessary for the number for their importance.
	: This {form} describes a controlled accelerating universe expansion where energy conditions are always satisfied, i.e., $P_{\phi}+\rho_{\phi} > 0$ \cite{steinhardt1,steinhardt2,steinhardt3,carroll1,quintessencecmbpeak,quintessenceholo,quintchakra2024,cosmofate,rollingscalarfield,steinhardt2024,wolf1,wolf2,wolf3}. {This usual DE form has been significantly studied in the literature in recent decades for the fascination it provokes and the realism of the models.}
	
	\item {\textbf{Phantom energy} $\alpha_Q {\bf <-1}$}: This {form} can usually describe an uncontrolled universe expansion accelerating toward a Big Rip event {(or singularity)} \cite{quintessencephantom,strongnegative,farnes,baumframpton,caldwell1,phantomdivide,phantomteleparallel1,ripphantomteleparallel2,phantomteleparallel3}. The energy condition is violated, i.e., $P_{\phi}+\rho_{\phi} \ngeq 0$. {But this DE form is fascinating because we can find new teleparallel solutions and physical models.}
	
	\item {\textbf{Cosmological constant} $\alpha_Q{\bf =-1}$}: This {primary DE form} is an intermediate limit between the {quintessence and phantom DE states}, where $P_{\phi}+\rho_{\phi} = 0$. A constant scalar field source $\phi=\phi_0$ {added by a positive scalar potential $V\left(\phi_0\right)>0$} will directly lead to this {primary DE state}. {Note that a non-positive scalar potential (i.e., $V\left(\phi_0\right)\leq 0$) will not lead to a positive cosmological constant and/or a DE solution.}

	\item {\textbf{Quintom models}}: This is a mixture of previous DE types, usually described by some double scalar field models \cite{quintom1,quintom2,quintom3,quintom4,quintomcoleytot,quintomteleparallel1}. This type of model is more complete to study and solve in general. Several types of models are in principle possible and these physical processes need further studies in the future.
\end{enumerate}

{The} %MDPI: We added indent here, please confirm. AL: OK
previous DE forms can be fundamentally defined in terms of the same scalar field with specific conditions. However the main interest of the current paper remains the new classes of teleparallel $F(T)$ solutions for static spherically symmetric spacetimes inside a scalar field, which may be useful for astrophysical purposes.

This paper is developed as follows. Section \ref{sect2} will summarize the teleparallel $F(T)$ gravity theory, the conservation laws, the used coframe/spin-connection pair and the static spherically symmetric FEs. Section \ref{sect3} will be for solving the unified FE in general and to find the teleparallel $F(T)$ solutions for power-law scalar field sources. Section \ref{sect4} will focus on exponential, logarithmic and other scalar field source teleparallel $F(T)$ solutions, {to which some graphical comparisons are added and a summary of the main results is proposed}. We will finish by the concluding remarks with future work recommendations in {Section} %MDPI: We moved the Notation part to  Abbreviations, please check and confirm. AL: It is OK.
\ref{sect5}.

\newpage

\section{Summary of Teleparallel Gravity and Field Equations}\label{sect2}

\subsection{Summary of Teleparallel Field Equations}

The teleparallel action integral is for any gravitational {source}%MDPI: 1. Please recheck all equations and make sure there are no duplicated equations in the whole manuscript. Thanks! 2. Please carefully check variable formatting (italic, bold, subscript, uppercase, etc.) throughout the manuscript to ensure the formatting is consistent and revise if needed. AL: No duplicated equations in the manuscript.
~\cite{Aldrovandi_Pereira2013,Bahamonde:2021gfp,Krssak:2018ywd,MCH,Coley:2019zld,SSpaper,nonvacSSpaper,nonvacKSpaper,roberthudsonSSpaper,scalarfieldKS,scalarfieldTRW}:
\begin{equation}\label{1000}
	S_{F(T)} = \int\,d^4\,x\,\left[\frac{h}{2\kappa}\,F(T)+\mathcal{L}_{Source}\right], 
\end{equation}
where $\kappa$ is the coupling constant and $h$ is the coframe determinant. By applying the least-action principle to Equation \eqref{1000}, we find the general FEs in terms of coframe and spin-connection~\cite{Aldrovandi_Pereira2013,Bahamonde:2021gfp,Krssak:2018ywd,MCH,Coley:2019zld,scalarfieldKS}:
\begin{align}\label{1000a}
	\kappa\,\Theta_a^{~~\mu} =&\, h^{-1}\,F_T\,\partial_{\nu}\left(h\,S_a^{~~\mu\nu}\right) + F_{TT}\,{S_a^{~~\mu\nu}\,\partial_{\nu}} T+\frac{F}{2}\,h_a^{~~\mu}-F_T\,\left(T^b_{~~a\nu}+\omega^b_{~~a\nu}\right)S_b^{~~\mu\nu} .
\end{align}
{{The} %MDPI: Please check if need to add indent here, if yes, please modify. Please check all uppercase letters after the equations. AL: Indentation OK.
	torsion tensor $T^a_{~~\mu\nu}$, the torsion scalar $T$ and the superpotential $S_a^{~~\mu\nu}$ are defined as follows \cite{Coley:2019zld}:
	\begin{align}
		T^a_{~~\mu\nu} =& \partial_{\mu}\,h^a_{~~\nu}-\partial_{\nu}\,h^a_{~~\mu}+\omega^a_{~~b\mu}h^b_{~~\nu}-\omega^a_{~~b\nu}h^b_{~~\mu},
		\\
		S_a^{~~\mu\nu}=& \frac{1}{2}\,\left(T_a^{~~\mu\nu}+T^{\nu\mu}_{~~a}-T^{\mu\nu}_{~~a}\right)-h_a^{~~\nu}\,T^{\lambda\mu}_{~~\lambda}+h_a^{~~\mu}\,T^{\lambda\nu}_{~~\lambda},
		\\
		T=&\frac{1}{2}\,T^a_{~~\mu\nu}S_a^{~~\mu\nu}.
	\end{align}	
}

{From} Equation~\eqref{1000a}, we find the symmetric and antisymmetric parts of FEs \cite{Coley:2019zld,SSpaper,nonvacSSpaper,nonvacKSpaper,roberthudsonSSpaper,scalarfieldKS,scalarfieldTRW}:
\begin{align}
	\kappa\,\Theta_{\left(ab\right)} =& F_T \overset{\ \circ}{G}_{ab}+F_{TT}\,S_{\left(ab\right)}^{\;\;\;\mu}\,\partial_{\mu} T+\frac{g_{ab}}{2}\,\left[F-T\,F_T\right],  \label{1001a}
	\\
	0 =& F_{TT}\left(T\right)\,S_{\left[ab\right]}^{\;\;\;\mu}\,\partial_{\mu} T. \label{1001b}
\end{align}
{The} canonical Energy--Momentum and its conservation laws (same as GR) are obtained from the $\mathcal{L}_{Source}$ term of Equation \eqref{1000} by the least-action principle and defined as \cite{Aldrovandi_Pereira2013,Bahamonde:2021gfp,nonvacSSpaper,nonvacKSpaper,roberthudsonSSpaper,scalarfieldKS,scalarfieldTRW,FTBcosmogholamilandry}
\begin{align}
	\Theta_a^{\;\;\mu}=&\frac{1}{h} \frac{\mathcal{L}_{Source}}{\delta h^a_{\;\;\mu}}, \quad \Rightarrow\quad \overset{\ \circ}{\nabla}\,_{\nu}\left(\Theta^{\mu\nu}\right)=0 .      \label{1001e}
\end{align}
{The} antisymmetric and symmetric parts of Equation \eqref{1001e} are \cite{Coley:2019zld,SSpaper,nonvacSSpaper,nonvacKSpaper,roberthudsonSSpaper,scalarfieldKS,scalarfieldTRW,FTBcosmogholamilandry}
\begin{equation}\label{1001c}
	\Theta_{[ab]}=0,\qquad \Theta_{(ab)}= T_{ab},
\end{equation}
{where $T_{ab}$ is the symmetric part, and then $\Theta_{ab}$ is purely symmetric.} Equation \eqref{1001c} is valid for {a source field interacting} with the metric {$g_{\mu\nu}$ associated with $h^a_{\;\;\mu}$ and $g_{ab}$}, and is not intricately coupled to the $F(T)$ gravity. {The covariant derivative $\overset{\ \circ}{\nabla}\,_{\nu}$ used in Equation \eqref{1001e} is defined in terms of GR with Levi--Civita connections for the current situation.} This consideration is valid only under the null hypermomentum condition defined by~\cite{golov3,nonvacSSpaper,nonvacKSpaper,scalarfieldKS,scalarfieldTRW}
\begin{align}\label{1001h}
	\mathfrak{T}_{ab}=\kappa\Theta_{ab}-F_T \overset{\ \circ}{G}_{ab}-F_{TT}\,S_{ab}^{\;\;\;\mu}\,\partial_{\mu} T-\frac{g_{ab}}{2}\,\left[F-T\,F_T\right]=0.
\end{align}
{{The} $\mathfrak{T}^{\mu\nu}\neq 0$ situations need} to satisfy more complex conservation law equations than Equation \eqref{1001e}, as shown in refs. \cite{hypermomentum1,hypermomentum2,hypermomentum3,golov3}.

\subsection{Static Spherically Symmetric Coframe and Spin-Connection Components}

In~the orthonormal gauge $g_{ab} =\eta_{ab}= Diag[-1,1,1,1]$, {any teleparallel geometry satisfies the relations}~\cite{olver1995equivalence}:
%\begin{subequations}
\begin{align}
	\mathcal{L}_{{\bf X}} {\bf h}^a =& \lambda^a_{~b} {\bf h}^b \quad\text{and}\quad  \mathcal{L}_{{\bf X}} {\omega}^a_{~bc} = 0, \label{Intro12}
\end{align}
%\end{subequations}
where ${\bf h}^a$ is the {differential form of} coframe, $\mathcal{L}_{{\bf X}}$ is the Lie derivative in terms of killing vectors (KVs) ${\bf X}$ and $\lambda^a_{~b}$ is the generator of  $\Lambda^a_{~b}$. {We must also satisfy for a pure teleparallel geometry the zero Riemann curvature} requirement ( refs.~\cite{Lucas_Obukhov_Pereira2009,Aldrovandi_Pereira2013,Bahamonde:2021gfp,MCH,Krssak:2018ywd,Coley:2019zld,Krssak_Pereira2015} and references therein):
\begin{align}\label{zerocurvatureeqsol}
	R^a_{~b\mu\nu}  =& \partial_{\mu}\omega^a_{~b\nu} -\partial_{\nu}\omega^a_{~b\mu}+\omega^a_{~e\mu}\omega^e_{~b\nu}-\omega^a_{~e\nu}\omega^e_{~b\mu} =0 ,
	\nonumber\\
	\Rightarrow\quad &\omega^a_{~b\mu} = \Lambda^a_{~c}\partial_{\mu}\Lambda_b^{~c}  .
\end{align} 
{The} solution of Equation~\eqref{zerocurvatureeqsol} leads to the teleparallel spin-connection defined in terms of $\Lambda^a_{~b}$. {We should note that $\omega^a_{~b\mu}=0$ for all proper frames and that $\omega^a_{~b\mu}\neq 0$ for all non-proper frames. }

The teleparallel spherically symmetric spacetimes {are defined, discussed and justified in} refs. \cite{SSpaper,nonvacSSpaper}. The static $r$-dependent spherically symmetric orthonormal coframe expression is defined by
\begin{equation}
	h^a_{~\mu} = \text{Diag}\left[ A_1(r),\, A_2(r),\,A_3(r),\,A_3(r) \sin(\theta)\right]. \label{VB:SS}
\end{equation}
Equation \eqref{VB:SS} is an invariant symmetry frame. The antisymmetric FE {solutions with a spin-connection defined under the form $\omega_{abc}=\omega_{abc}(\chi(r),\psi(r))$ are exactly} $\chi = n\,\pi$ and $\psi=0$, where $n \in \mathbb{Z}$ is an integer and $\cos\,\chi=\cos\left(n\,\pi\right)=\pm 1 = \delta$ \cite{SSpaper,nonvacSSpaper}. {The non-zero static $r$-dependent spin-connection components are}
\begin{align}\label{1050}
	\omega_{233} = \omega_{244} = \frac{\delta}{A_3(r)},~ \omega_{344} = - \frac{\cot(\theta)}{A_3(r)}. 
\end{align}
{{The} Equation \eqref{1050} terms are also similar to those obtained in refs. \cite{Krssak:2018ywd,golov3}. {Equations \eqref{VB:SS} and \eqref{1050}} recover the five arbitrary functions required for a teleparallel geometry: $A_1$, $A_2$, $A_3$, $\psi$ and $\chi$. We note that {Equations \eqref{VB:SS} and \eqref{1050}} are also solutions to Equations \eqref{Intro12} and \eqref{zerocurvatureeqsol}, and similar results were found for a spherically symmetric metric with a non-invariant proper frame approach \cite{pfeifer2,sharif2009teleparallel,hohmann2019modified,pfeifer2022quick}. }

\subsection{Static Scalar Field Energy-Momentum Source}

The scalar field source Lagrangian density $\mathcal{L}_{Source}$ is defined as \cite{Bahamonde:2021gfp,coleylandrygholami,scalar2,FTBcosmogholamilandry,scalarfieldKS}
\begin{align}\label{1002}
	\mathcal{L}_{Source} = \frac{h}{2}{\overset{\ \circ}{\nabla}}\,_{\nu}\phi\,\overset{\ \circ}{\nabla}\,^{\nu}\phi -h\,V\left(\phi\right),
\end{align}
where $V\left(\phi\right)$ is the scalar potential. The perfect fluid tensor $T_{ab}$ is defined as \cite{hawkingellis1,coleybook,scalar2,cosmofluidsbohmer}:
\begin{align}\label{1001d}
	T_{ab}= \left(P_{\phi}+\rho_{\phi}\right)\,u_a\,u_b+g_{ab}\,P_{\phi},
\end{align}
where $u_a=(-1,\,0,\,0,\,0)$ for a stationary fluid, $P_{\phi}$ and $\rho_{\phi}$ are the pressure and density equivalents in terms of scalar field defined by \cite{hawkingellis1,coleybook,scalar2,cosmofluidsbohmer}
\begin{align}\label{EoSscalar}
	P_{\phi} = \frac{\phi'^2}{2}-V(\phi) \quad\text{and}\quad \rho_{\phi} = \frac{\phi'^2}{2}+V(\phi),
\end{align}
{where $\phi=\phi(r)$.} The DE index (or quintessence index) $\alpha_Q$ {will be} \cite{steinhardt1,steinhardt2,steinhardt3,steinhardt2024,scalarfieldKS,wolf1,wolf2,wolf3}
\begin{align}\label{Quintessenceindex}
	\alpha_Q =& \frac{P_{\phi}}{\rho_{\phi}} = \frac{\phi'^2-2V\left(\phi\right)}{\phi'^2+2V\left(\phi\right)} {= -1+\frac{2\phi'^2}{\phi'^2+2V\left(\phi\right)}}.
\end{align}
{From} Equation \eqref{1001e} and by using Equation \eqref{EoSscalar}, the static conservation law equation is expressed by \cite{SSpaper,nonvacSSpaper}
\begin{align}
	&\frac{dV}{d\phi}=\phi'\,\left(\ln\,A_1\right)'+\phi''. \label{2201}
\end{align}
Equation \eqref{2201} will be useful to determine the scalar potential expression for any $A_1(r)$ coframe component and $\phi(r)$ definition.

\subsection{Static Scalar Field Source Field Equations}

From the FE components in Appendix \ref{appena}, the symmetric FEs and torsion scalar expressions are \cite{SSpaper,nonvacSSpaper}:
\begin{align}
	F_T=& F_T(0)\,\exp\left[\int_{r(T)}\,dr'\,\frac{g_1(r')}{k_1(r')}\right] , \label{2202a}
	\\
	\kappa\,\phi'^2 =& 2\,F_T\,\left[g_2(r)-\left(\frac{g_1(r)}{k_1(r)}\right)\,k_2(r)\right] , \label{2202b}
	\\
	\kappa\,\left[\phi'^2+2V(\phi(r))\right] =& -F + 4\,F_T\,\left[g_3(r)-\left(\frac{g_1(r)}{k_1(r)}\right)\,k_2(r)\right] , \label{2202c}
	\\
	T(r)=& -2\left(\frac{\delta}{A_3}+\frac{A_3'}{A_2\,A_3}\right)\left(\frac{\delta}{A_3}+\frac{A_3'}{A_2\,A_3}+\frac{2\,A_1'}{A_1\,A_2}\right), \label{2202d}
\end{align}
where $g_i$ and $k_i$ components are expressed by Equations \eqref{2922a}--\eqref{2922d} in Appendix \ref{appena} in terms of $A_1$, $A_2$, $A_3$ (and derivatives in $r$). However, we can simplify and merge the {Equations \eqref{2202a}--\eqref{2202c}} to obtain a potential dependent equation leading to $F(T)$ solutions in terms of a $r(T)$ relation from Equation \eqref{2202d}:
%\small
\vspace{-8pt}
%\begin{adjustwidth}{-\extralength}{0cm}
	%\centering %% If there is a figure in wide page, please release command \centering
	\begin{align}\label{2203}
		F(T) =& -2\kappa\,V(T) + 2\,F_T(0)\,\exp\left[\int_{r(T)}\,dr'\,\frac{g_1(r')}{k_1(r')}\right]\left[2\,g_3(r(T))-g_2(r(T))-\left(\frac{g_1(r(T))}{k_1(r(T))}\right)\,k_2(r(T))\right] ,
	\end{align}
%\end{adjustwidth}
\normalsize
where $V(T)=V(\phi(r(T)))$ is the scalar potential and $r(T)$ can be found by using Equation \eqref{2202d} {as a characteristic equation. We need to specify that any Equation \eqref{2203} teleparallel solution satisfies individually each of the Equations \eqref{2202a}--\eqref{2202c} by the uniqueness property of differential equation systems.} From Equation \eqref{2203}, we will be able to compute any $F(T)$ for any coframe ansatz, a potential $V(T)$ satisfying Equation \eqref{2201} and by using the $g_i$ and $k_i$ components.

%%newpage

\section{Power-Law Scalar Field Solutions}\label{sect3}

We will assume, {as done similarly in refs. \cite{baha10,baha4,scalarizedbhnew,sharifmodels1}}, a power-law scalar field defined by
\begin{align}\label{3000}
	\phi(T)=p_0\,\left[r(T)\right]^p,
\end{align}
{where $p$ and $p_0$ are $\mathbb{R}$-valued numbers.} For the most of the coming developments, we will use the power-law coframe component ansatz defined as
\begin{align}\label{powerlaw}
	A_1(r)=a_0\,r^a , \quad\quad\quad A_2(r)=b_0\,r^b . 
\end{align}
{By} assuming a power-law coframe ansatz, we will solve Equation \eqref{2201} to find the scalar potential $V(\phi)$ by using Equations \eqref{3000} and \eqref{powerlaw} for the following cases:
\begin{enumerate}
	\item \textbf{{General ($p\neq 1$):}}
	\begin{align}
		&\frac{dV}{d\phi}=p\,(a+p-1)\,p_0^{2/p}\,\phi^{1-2/p},
		\nonumber\\
		&\Rightarrow\,V(\phi)=V_0+\frac{p^2\,p_0^{2/p}\,(a+p-1)}{2(p-1)}\,\phi^{2-2/p}, \label{3001}
	\end{align}
	{{where $V_0>0$.} By substituting Equation \eqref{3000} into Equation \eqref{3001}, we find the $V(T)$ potential:}
	
	\begin{align}
		& V(T)=V_0+\frac{p^2\,p_0^{2}\,(a+p-1)}{2(p-1)}\,\left(r(T)\right)^{2p-2}, \label{3004}
	\end{align}
	where $r(T)$ is the characteristic equation solution defined from Equation \eqref{2202d}. {The Equation \eqref{Quintessenceindex} for dark energy index is
		\begin{align}\label{3007}
			\alpha_Q =&  -1+\frac{p^2\,p_0^{2}\,\left(r(T)\right)^{2p-2}}{\left(\frac{p^2\,p_0^{2}}{p-1}\right)\left(\frac{a}{2}+p-1\right)\,\left(r(T)\right)^{2p-2}+V_0}.
		\end{align}	
		{The} general power-law scalar field defined by Equation \eqref{3000} and by Equation \eqref{3001} can describe a tachyon model as shown in ref. \cite{sharifmodels1}. }
	
	\item {${\bf p=1}$:}
	\begin{align}
		&\frac{dV}{d\phi}=a\,p_0^{2}\,\phi^{-1}\quad\quad\Rightarrow\,V(\phi)=V_0+a\,p_0^2\,\ln\,\phi. \label{3002}
	\end{align}
	{By} substituting Equation \eqref{3000} into Equation \eqref{3002}, we find as $V(T)$:
	\begin{align}
		& V(T)=V_0+a\,p_0^2\,\ln\,p_0+a\,p_0^2\,\ln\,\left(r(T)\right), \label{3005}
	\end{align}
	where ${r(T)>0}$ is the characteristic equation solution defined from Equation \eqref{2202d}. {The Equation \eqref{Quintessenceindex} will be in this case:
		\begin{align}\label{3008}
			\alpha_Q =&  -1+\left[\frac{1}{2}+\frac{V_0}{p_0^2}+a\,\ln\,p_0+a\,\ln\,\left(r(T)\right)\right]^{-1}.
		\end{align}
	}
	
	\item {\textbf{Ordinary matter limit} (${\bf p\gg 1}$)}:
	\begin{align}
		&\frac{dV}{d\phi}=p^2\,\phi\quad\quad\Rightarrow\,V(\phi)=V_0+\frac{p^2}{2}\,\phi^{2}. \label{3003}
	\end{align}
	{By} substituting Equation \eqref{3000} into Equation \eqref{3003}, we find as $V(T)$:
	\begin{align}
		& V(T)=V_0+\frac{p^2\,p_0^2}{2}\,\left(r(T)\right)^{2p}, \label{3006}
	\end{align}
	where $r(T)$ is the characteristic equation solution defined from Equation \eqref{2202d}. {Equation~\eqref{3003}} is a simple harmonic oscillator (SHO) potential. This last case is also valid for any coframe ansatz, not only the power-law case. {Finally, Equation \eqref{Quintessenceindex} becomes under this limit:
		\begin{align}\label{3009}
			\alpha_Q \approx &  -1+\frac{\left(r(T)\right)^{2p}}{\left(r(T)\right)^{2p}+\frac{V_0}{p^2\,p_0^{2}}}\,\approx\,-1+1\,\rightarrow\,  0.
		\end{align}	
		Equation \eqref{3009} leads to the ordinary matter (or dust) equivalent EoS under the $|p|\,\rightarrow\,\infty$ limit (i.e., $P_{\phi}=0$ for any $\rho_{\phi}$ density of ordinary matter expression). This case describes an ordinary matter and/or a bosonic scalar field. For the current paper, we will consider this case as the \textbf{{ordinary matter limit}%MDPI: Please confirm if the bold is unnecessary and can be removed. The following highlights are the same. AL: Bold is necessary to highlight the importance of the Ordinary matter limit here.
		} because Equation \eqref{3009} under the $|p|\,\rightarrow\,\infty$ limit leads to $\alpha_Q=0$.
	}
\end{enumerate}

{{For} %MDPI: We added indent here, please check. AL: OK
	physical process purposes, the interpretation of Equations \eqref{3001}--\eqref{3008} can lead to the different forms of dark energy for the following reasons:
	\begin{enumerate}
		
		\item \textbf{{Quintessence:}} We must consider a value of $\alpha_Q$ higher than $-\frac{1}{3}$ because of dark energy's physical limit. Under the consideration of $-1<\alpha_Q<-\frac{1}{3}$ and from {Equations \eqref{3007}, \eqref{3008} and \eqref{3009}}'s respective results, we find the constraints to satisfy for the quintessence process {as} %MDPI: Please check if bold format in Equations (36--38) are unnecessary and can be removed. AL: Bold is necessary for clarifying the conditions.
		\begin{align}
			-\frac{2V_0}{p_0^2\,p^2\,\left(2+\frac{a}{p-1}\right)}  <  \left(r(T)\right)^{2p-2}\, < \, & \, \frac{2V_0}{p_0^2\,p^2\,\left(1-\frac{a}{p-1}\right)} &\, \textbf{($p\neq 1$)} ,\label{3007a}
			\\
			\frac{1}{p_0}\,\exp\left(\frac{1}{a}\,\left(1-\frac{V_0}{p_0^2}\right)\right)\,   < & \, r(T) \, & \, \textbf{($a>0$ and $p=1$)},    \label{3008a}
			\\
			\frac{1}{p_0}\,\exp\left(-\frac{1}{a}\left(\frac{1}{2}+\frac{V_0}{p_0^2}\right)\right)\,  < & \, r(T) \, &\, \textbf{($a<0$ and $p=1$)} .  \label{3008c}
		\end{align}
		{The} Equation \eqref{3009} condition for $|p|\,\rightarrow\,\infty$ leads to $r(T)\,\rightarrow\,0$ and this last case cannot lead to any DE scalar field. The constraint system of Equations \eqref{3007a}--\eqref{3008c} can yield, in principle, a significant number of possible solutions, because of $r(T)$ terms inside each of the equations. Therefore, Equations \eqref{3007a}--\eqref{3008c} provide a minimal value for a $r(T)$ characteristic equation solution and Equation \eqref{3007a} will lead to a maximal value for $r(T)$ and $p$.

		\item \textbf{{Phantom energy:}} Because of $\alpha_Q<-1$ in this type of model, we will find constraints from Equations \eqref{3007} and \eqref{3008} to {satisfy:} %MDPI: Please check if bold format in Equations (39--40) are unnecessary and can be removed. AL: Bold is necessary.
		\begin{align}
			\left(r(T)\right)^{2p-2} &<  -\frac{2V_0}{p_0^2\,p^2\,\left(2+\frac{a}{p-1}\right)}  \quad\quad & \textbf{($p\neq 1$)} , \label{3007b}
			\\
			r(T) &< \frac{1}{p_0}\,\exp\left(-\frac{1}{a}\left(\frac{1}{2}+\frac{V_0}{p_0^2}\right)\right)  \quad\quad & \textbf{($p=1$)}. \label{3008b}
		\end{align}
		Equation \eqref{3009}, under the $p\,\rightarrow\,\infty$ limit, cannot consistently lead to any phantom DE model. Once again, in the case of Equations \eqref{3007b} and \eqref{3008b}, there are a significant number of possible solutions for $r(T)$ and $p$. But the main point is that Equations \eqref{3007b} and  \eqref{3008b} provide, respectively, a minimal and a maximal value for $r(T)$.

	\end{enumerate}
}

\subsection{Power-Law Ansatz for $A_3=c_0$}\label{sect31}

By using Equation \eqref{powerlaw} and setting $A_3=c_0$, we find that the characteristic equation from the torsion scalar defined by Equation \eqref{2202d} is \cite{nonvacSSpaper}
\begin{align}\label{3010}
	0=&	\frac{c_0^2}{2}\,T+1+\frac{2\delta\,c_0}{b_0}\,r^{-(1+b)}\quad\quad\Rightarrow\,r^{-(1+b)}(T)=-\frac{\delta\,b_0}{2c_0}\left(1+\frac{c_0^2}{2}\,T\right){=-\frac{\delta\,b_0}{2c_0}\,u(T),} 
\end{align} 
{where $u(T)=1+\frac{c_0^2}{2}\,T \neq 0$, $b_0\neq 0$, $c_0\neq 0$ and $b\neq -1$.} In terms of Equation \eqref{3010}, we find~that
\begin{align}\label{3012}
	\phi(T) =\phi_0\,{\left(u(T)\right)}^{-p/(1+b)},
\end{align}
where $\phi_0=p_0\,\left(-\frac{2\delta\,c_0}{b_0}\right)^{p/(1+b)}$. By using Equation \eqref{2923} in Appendix \ref{appena}, Equation \eqref{2203} becomes
%\small
\vspace{-10pt}
%\begin{adjustwidth}{-\extralength}{0cm}
	%\centering %% If there is a figure in wide page, please release command \centering
	\begin{align}\label{3011}
		F(T) =& -2\kappa\,V(T) - \frac{2\,F_T(0)}{c_0^2}\frac{\left[2-a{\left(u(T)\right)}\right]^{\frac{(2a-b-1)}{(b+1)}}}{{\left(u(T)\right)}^{\frac{a}{(b+1)}}}\exp\left[\frac{2}{(b+1){u(T)}}\right]\left[\frac{\frac{a(1+a+b)}{4}\,{\left(u(T)\right)}^2-a{u(T)}-1}{\left[-\frac{a}{2}\,{u(T)}+1\right]}\right] ,
	\end{align}
%\end{adjustwidth}
\normalsize
where $b \neq -1$. The potential $V(T)$ in Equation \eqref{3011} from Equations \eqref{3001}--\eqref{3003} is for the following situations:
\begin{enumerate}
	\item \textbf{{General  ($p\neq 1$):}} %MDPI: Please check if the variable/equation should be also bold. AL: OK for keeping in bold.
	Equation \eqref{3001} becomes:
	\begin{align}
		V(T)=V_0+\frac{p^2\,p_0^{2}\,(a+p-1)}{2(p-1)}\,\left(-\frac{2\delta\,c_0}{b_0}\right)^{2(p-1)/(1+b)}\,{\left(u(T)\right)}^{-2(p-1)/(1+b)}. \label{3013}
	\end{align}
	
	\item {${\bf p=1}$}: Equation \eqref{3002} becomes:
	\begin{align}
		&V(T)=\tilde{V}_0-\frac{a\,p_0^2}{(1+b)}\,\ln\,{\left(u(T)\right)}, \label{3015}
	\end{align}
	where $\tilde{V}_0=V_0+a\,p_0^2\,\ln\,\phi_0$.
	
	\item {\textbf{Ordinary matter limit} (${\bf p\gg 1}$)}: Equation \eqref{3003} becomes:
	\begin{align}
		&V(T)=V_0+\frac{p^2\,p_0^2}{2}\,\left(\frac{2\,c_0}{b_0}\right)^{2p/(1+b)}\,{\left(u(T)\right)}^{-2p/(1+b)}. \label{3017}
	\end{align}

\end{enumerate}

%\frac{g_1}{k_1} = \frac{\left[\left(a(1-a+b)\right)\,r^{-2b-2}-\left(\frac{b_0}{c_0}\right)^2\right]}{\left[a\,r^{-2b-1}+\delta\,\left(\frac{b_0}{c_0}\right)\,r^{-b}\right]} ,\quad  g_2=0 , \quad g_3 = -\left(\frac{\delta\,a}{b_0\,c_0}\right)\,r^{-b-1} , 
%\quad k_2=k_3=\left(\frac{\delta}{b_0\,c_0}\right)\,r^{-b}

\subsection{Power-Law Ansatz for $A_3=r$}\label{sect32}

By using Equation \eqref{2202d} for $A_3=r$, we find as general characteristic equation \cite{nonvacSSpaper}:
\begin{align}\label{3100}
	0=& \frac{b_0^2\,T}{2}+b_0^2\,r^{-2}+2\delta\,b_0\,(1+a)\,r^{-2-b}+(2\,a+1)\,r^{-2-2b}.
\end{align}
Equation \eqref{2203}, in terms of Equation \eqref{2924} in Appendix \ref{appena}, is
\small
\begin{align}\label{3101}
	F(T) =& -2\kappa\,V(T) + \frac{2\,F_T(0)}{b_0^2}\,\exp\left[\int_{r(T)}\,dr'\,r'^{b-1}\frac{\left[{B_{ab}}\,r'^{-2b}-b_0^2\right]}{\left[(a+1)\,r'^{-b}+\delta\,b_0\right]}\right] \Bigg[(b-3a-2)\,r^{-2b-2}(T) 
	\nonumber\\
	&\;-2\delta\,b_0\,(a+1)\,r^{-b-2}(T) -\left(\frac{\left[{B_{ab}}\,r^{-2b}(T)-b_0^2\right]}{\left[(a+1)\,r^{-b}(T)+\delta\,b_0\right]}\right)\left(r^{-b}(T)+\delta\,b_0\right)\,r^{-2}(T)\Bigg] ,
\end{align}
\normalsize
{where $B_{ab}=2a-a^2+ab+b+1$.} Equation \eqref{3101} leads to several solutions depending on the $a$ and $b$ values satisfying Equation \eqref{3100}. The analytical cases leading to new teleparallel $F(T)$ solutions are {about the same ones as in ref. \cite{nonvacSSpaper}. This to more easily obtain the result comparisons.
	
	\subsubsection{Flat Cosmological Case {${\bf a=b=0}$}}\label{sect321}

	In this case,} Equation \eqref{3100} becomes:
\begin{align}
	0=& \frac{b_0^2\,T}{2}+\left(1+\delta\,b_0\right)^2\,r^{-2} \quad \Rightarrow\;r^{-2}(T)=\frac{b_0^2}{2\left(1+\delta\,b_0\right)^2} (-T)=C_{00}\,(-T), \label{3102}
\end{align}
{where $T\leq 0$ for a {$\mathbb{R}$-valued} $r(T)$} {and $b_0\neq \left\lbrace 0,\,-\delta \right\rbrace$.} The scalar field $\phi(T)$ will be from the definition of Equation \eqref{3000}:
\begin{align}\label{3103}
	\phi(T) =p_0\,\left(\frac{\sqrt{2}\left(1+\delta\,b_0\right)}{b_0}\right)^{p}\,(-T)^{-p/2}=\phi_{00}\,(-T)^{-p/2}.
\end{align}
{By} substituting Equations \eqref{3102} and \eqref{3103} into Equation \eqref{3101}, we find that
\begin{align}\label{3104}
	F(T) =& -2\kappa\,V(T) - \frac{2\,F_T(0)}{b_0^2}\frac{\left(3-\delta\,b_0\right)}{\left(1+\delta\,b_0\right)^{\delta\,b_0}}\,\left[\frac{b_0}{\sqrt{2}}\right]^{\left(1+\delta\,b_0\right)}\,(-T)^{\frac{\left(1+\delta\,b_0\right)}{2}} ,
\end{align}
where $V(T)$ is for the following cases:
\begin{enumerate}
	\item \textbf{{General ($p\neq 1$):}}%MDPI:  Please check if the variable/equation should be also bold. AL: OK for bold format.
	
	\begin{align}
		&V(T)=V_0+\frac{p^2\,p_0^{2}}{2}\,\left(\frac{\sqrt{2}\left(1+\delta\,b_0\right)}{b_0}\right)^{2p-2}\,(-T)^{1-p}. \label{3105}
	\end{align}
	
	\item {${\bf p=1}$}: $V(\phi(T))=V_0=$ constant. Equation \eqref{3104} will be a power-law like $F(T)$ solution.
	
	\item {\textbf{Ordinary matter limit} (${\bf p\gg 1}$)}:
	\begin{align}
		&V(T)=V_0+\frac{p^2\,p_0^{2}}{2}\,\left(\frac{\sqrt{2}\left(1+\delta\,b_0\right)}{b_0}\right)^{2p}\,(-T)^{-p}. \label{3106}
	\end{align}
\end{enumerate}
{{The} last case, described by Equation \eqref{3104} for the $F(T)$ solution with Equations \eqref{3105} and \eqref{3106} as possible $V(T)$ potentials, goes to the same direction as ref. \cite{sharifmodels1}'s studied solutions. However, ref. \cite{sharifmodels1}'s studied model intrinsically assumes a power-law teleparallel $F(T)$ solution similar to Equations \eqref{3105}--\eqref{3106} without taking into account the homogeneous part of the $F(T)$ solution. Equations \eqref{3104}--\eqref{3106} go further than ref. \cite{sharifmodels1}'s studied solutions by including the homogeneous parts of the solutions: this constitutes a good progress.}

If we set $b_0=\delta$, Equation \eqref{3103} simplifies as
\begin{align}\label{3107}
	F(T) =& -2\kappa\,V(T) + F_T(0)\,T ,
\end{align}
where $V(T)$ is:
\begin{enumerate}
	\item \textbf{{General ($p\neq 1$):}}%MDPI:  Please check if the variable/equation should be also bold. AL: OK for bold format.
	
	\begin{align}
		&V(\phi(T))=V_0+p^2\,p_0^{2}\,\left(2\right)^{3p-4}\,(-T)^{1-p}. \label{3108}
	\end{align}
	
	\item {${\bf p=1}$}: $V(\phi(T))=V_0=$ constant. Equation \eqref{3107} will be a GR (TEGR-like) solution.
	
	\item {\textbf{Ordinary matter limit} (${\bf p\gg 1}$)}:
	\begin{align}
		&V(T)=V_0+p^2\,p_0^{2}\,\left(2\right)^{3p-1}\,(-T)^{-p}. \label{3109}
	\end{align}
\end{enumerate}
{	
	
	\subsubsection{General {${\bf a\neq \left\lbrace -1,\, -\frac{1}{2} \right\rbrace}$} Cases}\label{sect322}

	There are four cases of $b$-parameter value leading to analytical $F(T)$ solutions according to Equation \eqref{3100}: }
\begin{enumerate}
	\item {${\bf b=0}$}: Equation \eqref{3100} becomes:
	\begin{align}\label{3112}
		0=& \frac{b_0^2\,T}{2}+\left(b_0^2+2\delta\,b_0\,(1+a)+2\,a+1\right)\,r^{-2},
		\nonumber\\
		&\quad\Rightarrow\,r^{-2}(T)=\frac{b_0^2}{2\,\left(b_0^2+2\delta\,b_0\,(1+a)+2\,a+1\right)}\,\left(-T\right)=C_{a0}\,\left(-T\right),
	\end{align}
	where $T \leq 0$ {and $b_0 \neq \left\lbrace 0,\,-\delta \right\rbrace $}. Equation \eqref{3000} for the scalar field is
	\begin{align}\label{3113}
		\phi(T)=p_0\,\left(\frac{2^{p/2}\,\left(b_0^2+2\delta\,b_0\,(1+a)+2\,a+1\right)^{p/2}}{b_0^{p}}\right)\,\left(-T\right)^{-p/2}=p_0\,\phi_{a0}\,\left(-T\right)^{-p/2}.
	\end{align}
	By substituting Equations \eqref{3112} and \eqref{3113} into Equation \eqref{3101}, we find that
	\small
	\begin{align}\label{3114}
		F(T) =& -2\kappa\,V(T) + \frac{2\,F_T(0)}{b_0^2}\,C_{a0}^{1-\frac{\left[\left(2a-a^2+1\right)-b_0^2\right]}{2\left[(a+1)+\delta\,b_0\right]}}\Bigg[-(3a+2)-2\delta\,b_0\,(a+1)
		\nonumber\\
		&\quad  -\left(\frac{\left[\left(2a-a^2+1\right)-b_0^2\right]}{\left[(a+1)+\delta\,b_0\right]}\right)\left(1+\delta\,b_0\right)\Bigg]\left(-T\right)^{1-\frac{\left[\left(2a-a^2+1\right)-b_0^2\right]}{2\left[(a+1)+\delta\,b_0\right]}},
	\end{align}
	\normalsize
	where the possible $V(T)$ are:
	\begin{enumerate}
		\item \textbf{{General ($p\neq 1$):}}%MDPI: Please check if the variable/equation should be also bold. AL: OK for bold format.
		
		\begin{align}
			&V(T)=V_0+\frac{p^2\,p_0^{2}\,(a+p-1)}{2(p-1)}\,\phi_{a0}^{2-2/p}\,\left(-T\right)^{1-p}. \label{3115}
		\end{align}
		
		\item {${\bf p=1}$}:
		\begin{align}
			&V(T)=V_0+a\,p_0^2\,\ln\,\left[p_0\,\phi_{a0}\right]-\frac{a\,p_0^2}{2}\,\ln\,\left(-T\right), \label{3116}
		\end{align}
		{where $T<0$ and $b_0 \neq \left\lbrace 0,\,-\delta \right\rbrace $.}
		
		\item {\textbf{Ordinary matter limit} (${\bf p\gg 1}$)}:
		\begin{align}
			&V(T)=V_0+\frac{p^2\,p_0^{2}}{2}\,\phi_{a0}^2\,\left(-T\right)^{-p}. \label{3117}
		\end{align}
		
	\end{enumerate}
	{{Here} again in this situation, Equations \eqref{3114}--\eqref{3117} are similar to ref. \cite{sharifmodels1}'s studied models, but we take into account the homogeneous part of the $F(T)$ solution. This case also improves ref. \cite{sharifmodels1}'s studied solutions and we will then progress by the same manner with more complex subcases below.}
	
	%%newpage
	
	\item {${\bf b=1}$}: Equation \eqref{3100} will be
	\begin{align}\label{3122}
		0=& \frac{b_0^2\,T}{2}+b_0^2\,r^{-2}+2\delta\,b_0\,(1+a)\,r^{-3}+(2\,a+1)\,r^{-4},
	\end{align}
	{The} solutions are
	% from here: simplify expression
	{\small
		\begin{align}
			r^{-1}(T)=&\;-\frac{\delta_1\,b_0(a+1)}{2(2a+1)}-\frac{\delta_1}{2}\Bigg[\Bigg(-\frac{2b_0^2}{3(2a+1)}+\frac{b_0^2(a+1)^2}{(2a+1)^2}+\frac{\sqrt[3]{w(T)}}{6\sqrt[3]{2}(2a+1)} 
			\nonumber\\
			&\;+\frac{2^{4/3}b_0^2\left(b_0^2+6\,(2a+1)\,T\right)}{3(2a+1)\sqrt[3]{w(T)}}\Bigg)\Bigg]^{1/2}-\frac{\delta_2}{2}\Bigg[-\frac{4b_0^2}{3(2a+1)}+\frac{2b_0^2(a+1)^2}{(2a+1)^2}-\frac{\sqrt[3]{w(T)}}{6\sqrt[3]{2}(2a+1)}
			\nonumber\\
			&\;+\frac{2^{4/3}b_0^2\left(b_0^2+6\,(2a+1)\,T\right)}{3(2a+1)\sqrt[3]{w(T)}}+\delta_1\,\frac{2\delta\,b_0^3\,a^2\,(a+1)}{(2a+1)^3}\Bigg[-\frac{2b_0^2}{3(2a+1)}+\frac{b_0^2(a+1)^2}{(2a+1)^2}
			\nonumber\\
			&\;+\frac{\sqrt[3]{w(T)}}{6\sqrt[3]{2}(2a+1)}+\frac{2^{4/3}b_0^2\left(b_0^2+6\,(2a+1)\,T\right)}{3(2a+1)\sqrt[3]{w(T)}}\Bigg]^{-1/2}\Bigg]^{1/2}, \label{3123}
		\end{align}
		\normalsize
		where $\left(\delta_1,\,\delta_2\right)=\left(\pm 1,\,\pm 1\right)$, $r(T)\neq 0$, $a\neq -\frac{1}{2}$,
		\vspace{-12pt}
%		\begin{adjustwidth}{-\extralength}{0cm}
			%\centering %% If there is a figure in wide page, please release command \centering
			\begin{align}
				w(T)=& 16b_0^6+\left(432(a+1)^2-288\,(2a+1)\right)\,b_0^4\,T 
				\nonumber\\
				&\,+\sqrt{\left(\left(16b_0^6+\left(432(a+1)^2-288(2a+1)\right)\,b_0^4\,T\right)^2-4b_0^6\left(4b_0^2+24(2a+1)\,T\right)^3\right)},
			\end{align}
%		\end{adjustwidth}
		and $w(T) \neq 0$.}  By substituting Equations \eqref{3000} and \eqref{3123} into Equation \eqref{3101}, we find~that
	\small
	\begin{align}\label{3125}
		F(T) =& -2\kappa\,V(T) - \frac{2\,F_T(0)}{b_0^2}\,\exp\left[-\delta\,b_0\,r(T)\right]\,\left[a+1+\delta\,b_0\,r(T)\right]^{\frac{2a^2-a-1}{a+1}}\left[r(T)\right]^{\frac{-a^2+3a+2}{a+1}-2} 
		\nonumber\\
		&\quad\times\, \Bigg[\frac{(3a+1)}{r^{2}(T)}+\frac{2\delta\,b_0\,(a+1)}{r(T)} +\frac{\left[\left(3a-a^2+2\right)\,r^{-2}(T)-b_0^2\right]}{\left[(a+1)\,r^{-1}(T)+\delta\,b_0\right]}\left(r^{-1}(T)+\delta\,b_0\right)\Bigg] ,
	\end{align}
	\normalsize
	where $V(T)$ are described by Equations \eqref{3004}, \eqref{3005} and \eqref{3006} with $r(T)$ defined by Equation \eqref{3123}.

	\item {${\bf b=-1}$}: Equation \eqref{3100} will be
	\begin{align}\label{3132}
		0=& \left(\frac{b_0^2\,T}{2}+2\,a+1\right)+2\delta\,b_0\,(1+a)\,r^{-1}+b_0^2\,r^{-2},
		\nonumber\\
		&\;\Rightarrow\,r^{-1}(T)=-\frac{\delta\,(a+1)}{b_0}+\delta_1 \sqrt{\frac{a^2}{b_0^2}-\frac{T}{2}} ,
	\end{align}
	where $\delta_1=\pm 1$, {$b_0\neq 0$ and $\frac{2a^2}{b_0^2}\geq T$.}  Equation \eqref{3000} is	
	\begin{align}\label{3133}
		\phi(r)=p_0\,\left(-\frac{\delta\,(a+1)}{b_0}+\delta_1 \sqrt{\frac{a^2}{b_0^2}-\frac{T}{2}}\right)^{-p},
	\end{align}
	By substituting Equations \eqref{3132} and \eqref{3133} into Equation \eqref{3101}, we find that
	\small
	\begin{align}\label{3134}
		F(T) =& -2\kappa\,V(T) - \frac{2\,F_T(0)}{b_0^2}\,\exp\left[\frac{\delta\,b_0}{r(T)}\right]\,\frac{\left[r(T)\right]^{a+1}}{\left[(a+1)\,r(T)+\delta\,b_0\right]^{-\frac{2a^2+a+1}{a+1}}}
		\nonumber\\
		&\quad\times\, \Bigg[3(a+1)+\frac{2\delta\,b_0\,(a+1)}{r(T)} +\frac{\left[a\left(1-a\right)\,r^{2}(T)-b_0^2\right]}{\left[(a+1)\,r(T)+\delta\,b_0\right]}\frac{\left(r(T)+\delta\,b_0\right)}{r^{2}(T)}\Bigg] ,
	\end{align}
	\normalsize
	{where $V(T)$ are described by Equations \eqref{3004}, \eqref{3005} and \eqref{3006} with $r(T)$ defined by Equation \eqref{3132}.}

	\item {${\bf b=-2}$}: Equation \eqref{3100} will be
	%	\small
%	\vspace{-6pt}
%	\vspace{-6pt}
%	\begin{adjustwidth}{-\extralength}{0cm}
		%\centering %% If there is a figure in wide page, please release command \centering
		\begin{align}\label{3142}
			0=& b_0^2+\left(\frac{b_0^2\,T}{2}+2\delta\,b_0\,(1+a)\right)\,r^2+(2\,a+1)\,r^{4},
			\nonumber\\
			&\;\Rightarrow\,r(T)=\frac{1}{2\sqrt{2a+1}}\,\sqrt{\delta_2\,b_0\sqrt{16a^2+8\delta\,(a+1)\,b_0\,T+b_0^2\,T^2} -4\delta\,b_0(a+1)-b_0^2\,T},
		\end{align}
%	\end{adjustwidth}
	%\normalsize
	{where $a\neq -\frac{1}{2}$.} Then, we will set the positive $r(T)$ case and $\delta_2=\pm 1$. The scalar field $\phi(T)$ is still described by Equation \eqref{3000} and then, by substituting {Equations \eqref{3000} and \eqref{3142}} into Equation \eqref{3101}, we find:
	\small
	\begin{align}\label{3143}
		F(T) =& -2\kappa\,V(T) - \frac{2\,F_T(0)}{b_0^2}\,\exp\left[\frac{\delta\,b_0}{2\,r^2(T)}\right] \frac{\left[r(T)\right]^{a+1}}{\left[(a+1)\,r^{2}(T)+\delta\,b_0\right]^{\frac{a^2+a+1}{a+1}}} 
		\nonumber\\
		&\quad\times\,\Bigg[(3a+4)\,r^{2}(T) +2\delta\,b_0\,(a+1) -\frac{\left[\left(a^2+1\right)\,r^{4}(T)+b_0^2\right]}{\left[(a+1)\,r^{2}(T)+\delta\,b_0\right]}\frac{\left(r^{2}(T)+\delta\,b_0\right)}{r^{2}(T)}\Bigg] ,
	\end{align}
	\normalsize
	where $V(T)$ are defined by Equations \eqref{3004}, \eqref{3005} and \eqref{3006} forms with $r(T)$ described by Equation \eqref{3142}.
	
\end{enumerate}
{
	\subsubsection{{${\bf a=-1}$} Cases}\label{sect323}	
	
}
\begin{enumerate}	
	{\item {${\bf b=0}$}: By setting $a=-1$, the} Equation \eqref{3100} becomes
	\begin{align}
		0=& \frac{b_0^2\,T}{2}+\left(b_0^2-1\right)\,r^{-2}\,\quad\Rightarrow\;	r^{-2}(T) = \left(\frac{b_0^2}{2\left(1-b_0^2\right)}\right)\,T,  \label{3152}
	\end{align}
	{where $T\neq0$ and $b_0\neq \left\lbrace 0,\,\delta\right\rbrace$.} The scalar field defined by Equation \eqref{3000} is
	\begin{align}\label{3153}
		\phi(T)=p_0\,\left(\frac{2^{p/2}\left(1-b_0^2\right)^{p/2}}{b_0^p}\right)\,T^{-p/2}=p_0\,\phi_{-1,0}\,T^{-p/2}.
	\end{align}
	{By} substituting Equations \eqref{3152} and \eqref{3153} into Equation \eqref{3101}, we find that
	%	\small
	\vspace{-8pt}	
%	\begin{adjustwidth}{-\extralength}{0cm}
		%\centering %% If there is a figure in wide page, please release command \centering
		\begin{align}\label{3154}
			F(T) =& -2\kappa\,V(T) + \frac{2\,F_T(0)}{b_0^3}\left(\frac{b_0^2}{2\left(1-b_0^2\right)}\right)^{1+\frac{\delta}{2b_0}\,\left(2+b_0^2\right)}\left[b_0+\delta\,\left(2+b_0^2\right)\left(1+\delta\,b_0\right)\right]\,T^{1+\frac{\delta}{2b_0}\,\left(2+b_0^2\right)} ,
		\end{align}
%	\end{adjustwidth}
	%\normalsize
	where $V(T)$ becomes for the cases:
	\begin{enumerate}
		\item \textbf{{General ($p\neq 1$):}}%MDPI: Please check if the variable/equation should be also bold. Same for the following AL: OK for bold format.
		
		\begin{align}
			&V(T)=V_0+\frac{p^2\,p_0^{2}\,(p-2)}{2(p-1)}\,\phi_{-1,0}^{2-2/p}\,T^{1-p}. \label{3155}
		\end{align}
		
		\item {${\bf p=1}$}:
		\begin{align}
			&V(T)=V_0-\,p_0^2\,\ln\,\left(p_0\,\phi_{-1,0}\right)+\frac{p_0^2}{2}\,\ln\,\left(T\right), \label{3156}
		\end{align}
		{where $T>0$.}
		
		\item {\textbf{Ordinary matter limit} (${\bf p\gg 1}$)}: 
		\begin{align}
			&V(T)=V_0+\frac{p^2\,p_0^{2}}{2}\,\phi_{-1,0}^{2}\,T^{-p}. \label{3157}
		\end{align}
	\end{enumerate}

	\item {{${\bf b\neq 0}$} cases: For $a=-1$,} the Equation \eqref{3101} simplifies as
	\small
	\begin{align}\label{3158}
		F(T) =& -2\kappa\,V(T) + \frac{2\,F_T(0)}{b_0^3}\,{ \exp\left[\frac{2\delta}{b_0\,b}\,\left[r(T)\right]^{-b}-\frac{\delta\,b_0}{b}\,\left[r(T)\right]^{b}\right]\Bigg[(b+3)\,b_0\,r^{-2b}(T)+2\delta\,r^{-3b}(T)}
		\nonumber\\
		&\quad {+\delta\,b_0^2\,\,r^{-b}(T) +b_0^3\Bigg]\,r^{-2}(T) ,}
		\nonumber\\
		=& {-2\kappa\,V(T) +\frac{2\,F_T(0)}{b_0^3}\,G_{-1,b}\left(r(T)\right)\,r^{-2}(T) ,   }
	\end{align}
	\normalsize
	where
	\vspace{-10pt}
%	\begin{adjustwidth}{-\extralength}{0cm}
		%\centering %% If there is a figure in wide page, please release command \centering
		\begin{align}\label{3158function}
			{G_{-1,b}\left(r(T)\right) = \,\exp\left[\frac{2\delta}{b_0\,b}\,\left[r(T)\right]^{-b}-\frac{\delta\,b_0}{b}\,\left[r(T)\right]^{b}\right]\Bigg[\frac{(b+3)\,b_0}{r^{2b}(T)}+\frac{2\delta}{r^{3b}(T)} +\frac{\delta\,b_0^2}{r^{b}(T)} +b_0^3\Bigg] ,}
		\end{align}
%	\end{adjustwidth}
	and the $V(T)$ expressions under Equation \eqref{3000}'s definition:
	\begin{enumerate}
		\item \textbf{{General ($p\neq 1$):}} Equation \eqref{3004} becomes
		\begin{align}
			& V(T)=V_0+\frac{p^2\,p_0^{2}\,(p-2)}{2(p-1)}\,\left[r(T)\right]^{2p-2}. \label{3159}
		\end{align}
		
		\item {${\bf p=1}$}: Equation \eqref{3005} becomes
		\begin{align}
			& V(T)=V_0-{p_0^2\,\ln\,\left[p_0\,r(T)\right],} \label{3160}
		\end{align}
		{where $p_0\,r(T)>0$.}
		
		\item {\textbf{Ordinary matter limit} (${\bf p\gg 1}$)}: Equation \eqref{3006} becomes
		\begin{align}
			& V(T)=V_0+\frac{p^2\,p_0^{2}}{2}\,\left[r(T)\right]^{2p}. \label{3161}
		\end{align}
	\end{enumerate}
\end{enumerate}	
{{All} the $r(T)$ solutions of Equation \eqref{3100} are presented in Table \ref{tableb1} in Appendix \ref{appenb}. The values of Equation \eqref{3158function}'s special function computation for any ${\bf b\neq 0}$ cases are presented in Table~\ref{tableb2} in Appendix \ref{appenb}. The $a=-1$ cases for $b\neq 0$ are summarized by Equations \eqref{3158}--\eqref{3161} and Tables \ref{tableb1} and \ref{tableb2} in Appendix \ref{appenb}. The {Table} %MDPI: We added Table here, please confirm. AL: OK for this change.
	\ref{tableb1} values of $r(T)$ must satisfy {Equations \eqref{3007a}--\eqref{3008c}} for a quintessence DE scalar field and Equations \eqref{3007b} and \eqref{3008b} for a phantom DE scalar field, as for the solutions in Sections \ref{sect321} and \ref{sect322}.}

\subsubsection{{${\bf a=-\frac{1}{2}}$} Cases}\label{sect324}	 

\begin{enumerate}
	
	\item {{${\bf b=0}$} case: By setting $a=-\frac{1}{2}$, the} Equation \eqref{3100} becomes
	\begin{align}\label{3242}
		0=& \frac{b_0\,T}{2}+\left(b_0+\delta\right)\,r^{-2} \quad \Rightarrow\;r^{-2}(T)=-{\frac{b_0}{2\left(b_0+\delta\right)}} \,T ,
	\end{align}
	{where $T\leq 0$ and $b_0\neq \left\lbrace0,\,-\delta \right\rbrace$.} Equation \eqref{3101} becomes by substituting {Equation~\eqref{3242}}
	\small
	\begin{align}\label{3243}
		F(T) =& -2\kappa\,V(T) {- \frac{2\delta\,F_T(0)}{b_0} \frac{\left(\frac{3}{4}-b_0^2\right)}{\left(\frac{1}{2}+\delta\,b_0\right)}\left[{\frac{b_0}{2\left(b_0+\delta\right)}}\right]^{\frac{\left(\frac{1}{4}+b_0^2\right)}{\left(1+2\delta\,b_0\right)}+1}\,(-T)^{\frac{\left(\frac{1}{4}+b_0^2\right)}{\left(1+2\delta\,b_0\right)}+1} ,}
	\end{align}
	\normalsize
	where $V(T)$ are defined in terms of Equation \eqref{3242} by
	\begin{enumerate}
		\item \textbf{{General ($p\neq 1$):}} Equation \eqref{3004} becomes
		\begin{align}
			& V(T)=V_0+\frac{p^2\,p_0^{2}\,\left(p-\frac{3}{2}\right)}{2(p-1)}\,\left({\frac{2\left(b_0+\delta\right)}{b_0}}\right)^{p-1}\,(-T)^{1-p}. \label{3244}
		\end{align}
		
		\item {${\bf p=1}$}: Equation \eqref{3005} becomes
		\begin{align}
			& V(T)=V_0+a\,p_0^2\,\ln\,p_0+\frac{p_0^2}{4}\,\ln\left({\frac{b_0}{2\left(b_0+\delta\right)}}\right)+\frac{p_0^2}{4}\,\ln\left(-T\right). \label{3245}
		\end{align}

		\item {\textbf{Ordinary matter limit} (${\bf p\gg 1}$)}: Equation \eqref{3006} becomes
		\begin{align}
			& V(T)=V_0+\frac{p^2\,p_0^2}{2}\,\left({\frac{2\left(b_0+\delta\right)}{b_0}}\right)^{p}\,(-T)^{-p}. \label{3246}
		\end{align}
		
	\end{enumerate}
	
	\item {${\bf b\neq 0}$} {cases: By setting $a=-\frac{1}{2}$, the} Equation \eqref{3101} will simplify as
	\small
	\begin{align}\label{3247}
		F(T) =& -2\kappa\,V(T) + \frac{2\,F_T(0)}{b_0^2}\, {\exp\left[\int_{r(T)}\,dr'\,r'^{b-1}\frac{\left[\frac{(2b-1)}{4}\,r'^{-2b}-b_0^2\right]}{\left[\frac{1}{2}\,r'^{-b}+\delta\,b_0\right]}\right]\Bigg[\left(b-\frac{1}{2}\right)\,r^{-2b}(T) }
		\nonumber\\
		&\quad {-\delta\,b_0\,r^{-b}(T) -\frac{\left[\frac{(2b-1)}{4}\,r^{-2b}(T)-b_0^2\right]}{\left[\frac{1}{2}\,r^{-b}(T)+\delta\,b_0\right]}\left(r^{-b}(T)+\delta\,b_0\right)\Bigg]\,r^{-2}(T) },
		\nonumber\\
		=& {-2\kappa\,V(T) + \frac{2\,F_T(0)}{b_0^2}\,G_{-\frac{1}{2},b}\left(r(T)\right)\,r^{-2}(T),}		
	\end{align}
	\normalsize
	where 
	\begin{align}\label{3247function}
		{G_{-\frac{1}{2},b}\left(r(T)\right)}= & {\, \exp\left[\int_{r(T)}\,dr'\,r'^{b-1}\frac{\left[\frac{(2b-1)}{4}\,r'^{-2b}-b_0^2\right]}{\left[\frac{1}{2}\,r'^{-b}+\delta\,b_0\right]}\right]\Bigg[\left(b-\frac{1}{2}\right)\,r^{-2b}(T) }
		\nonumber\\
		&\quad {-\delta\,b_0\,r^{-b}(T) -\frac{\left[\frac{(2b-1)}{4}\,r^{-2b}(T)-b_0^2\right]}{\left[\frac{1}{2}\,r^{-b}(T)+\delta\,b_0\right]}\left(r^{-b}(T)+\delta\,b_0\right)\Bigg] , }
	\end{align}
	and $V(T)$ are defined by
	\begin{enumerate}
		\item \textbf{{General ($p\neq 1$):}} Equation \eqref{3004} becomes
		\begin{align}
			& V(T)=V_0+\frac{p^2\,p_0^{2}\,\left(p-\frac{3}{2}\right)}{2(p-1)}\,\left[r(T)\right]^{2p-2}. \label{3248}
		\end{align}
		
		\item {${\bf p=1}$}: Equation \eqref{3005} becomes
		\begin{align}
			& V(T)=V_0-{\frac{p_0^2}{2}\,\ln\,\left[p_0\,r(T)\right]}, \label{3249}
		\end{align}
		{where $p_0\,r(T)>0$.}
		
		\item {\textbf{Ordinary matter limit} (${\bf p\gg 1}$)}: Equation \eqref{3006} keeps the same form.
	\end{enumerate}
\end{enumerate}

{All the $r(T)$ solutions of Equation \eqref{3100} are presented in Table \ref{tableb3} in Appendix \ref{appenb}. The values of Equation \eqref{3247function}'s special function computation for any ${\bf b\neq 0}$ cases are presented in Table \ref{tableb4} in Appendix \ref{appenb}. The $a=-\frac{1}{2}$ cases for $b\neq 0$ are summarized by {Equations \eqref{3247}--\eqref{3249}} and Tables \ref{tableb3} and \ref{tableb4} in Appendix \ref{appenb}. The {Table} %MDPI: We added ``'Table'' here, please confirm. AL: OK for this change.
	\ref{tableb3} values of $r(T)$ must again satisfy Equations \eqref{3007a}--\eqref{3008c} for a quintessence DE scalar field and {Equations \eqref{3007b} and \eqref{3008b}} for a phantom DE scalar field, as for the solutions in Sections \ref{sect321}--\ref{sect323}.}

% 	&\frac{g_1}{k_1} = \frac{\left[{ B_{ab}}\,r^{-2b}-b_0^2\right]}{\left[(a+1)\,r^{2(1-b)-1}+\delta\,b_0\,r^{(1-b)}\right]} ,~
% & g_2 = \frac{(a+b)}{b_0^2}\,r^{-2b-2} , &
% \nonumber\\
% & g_3 = \frac{1}{b_0^2}\Bigg[(-a+b-1)\,r^{-2b-2}-\delta\,b_0\,(a+1)\,r^{-b-2}\Bigg] ,~ & k_2= k_3 = \frac{1}{b_0^2}\Bigg[r^{-2b-1}+\delta\,b_0\,r^{-b-1}\Bigg] . &

\subsection{Other Ansatzes and Possible Comparisons with the Literature}\label{sect33}

The most of Sections \ref{sect31} and \ref{sect32} new teleparallel $F(T)$ solutions were computed from the general Equation \eqref{2203} leading to Equations \eqref{3011} and \eqref{3101} for $A_3=c_0$ and $A_3=r$ classes of solutions, respectively. These new solutions can be easily and directly comparable to the linear perfect fluids teleparallel $F(T)$ solutions found in ref. \cite{nonvacSSpaper} by a very similar solving approach. The main advantage in favor of this paper concerning the scalar field sources is the general teleparallel $F(T)$ solution-computing formula stated by Equation~\eqref{2203}. This general easy-to-use $F(T)$ computation formula can be used for any scalar field and any coframe component ansatzes. In more practical and technical words, we can set in principle any $A_1$ and $A_2$ component ansatz for either $A_3=c_0$ or $A_3=r$ solutions and generate any new teleparallel $F(T)$ solution depending on the ansatz parameter. A very similar approach was used to find the teleparallel $F(T)$ solutions for general Kantowski--Sachs (KS) spacetimes (pure time-dependent spherically spacetimes) and for teleparallel Robertson--Walker (TRW) $F(T)$ solutions with scalar field sources in both cases \cite{scalarfieldKS,scalarfieldTRW}. We have found in the two cases those easy-to-compute teleparallel $F(T)$ solution formula to generate the new solutions applicable to cosmological models, especially the dominating DE universe cases. However, the current paper confirms the relevance of using the scalar field sources, but for focusing on $r$-coordinate-based astrophysical system applications. {There are some typical examples in the recent literature and we should also propose to compare our results with observational data.

	A first example from the literature are the Born--Infeld BH solutions in teleparallel gravity. In the first case, several of the previous solutions (especially for larger $b$-parameters) can be related to a kind of Born--Infeld BH solutions. This type of solution is defined in the {literature as} %MDPI: We reordered references so that their first citation appears in numerical order. Please confirm. AL: OK for this change.
	\cite{baha6,borninfeldadd}
	\begin{align}\label{borninfeldFT}
		F(T) = T_0\,\left(\sqrt{1+\frac{2T}{T_0}}-1\right) \, \approx \,T+\frac{T^2}{T_0} \,\sum_{k=0}^{\infty}\,a_k\,\left(\frac{T}{T_0}\right)^k,
	\end{align}
	where the last term is valid for $\frac{T}{T_0}\ll 1$. This last expression shows that a Born--Infeld solution can be defined as a sum of power-law terms. For $a_k=\frac{1}{k!}$, Equation \eqref{borninfeldFT} will be expressed as $F(T)=T+\frac{T^2}{T_0}\,\exp\left(\frac{T}{T_0}\right)$. Some of the previous new $F(T)$ solutions are close to those of Born--Infeld BH solutions, especially higher values of $b$-parameters where the geometry terms are dominating compared to the $V(T)$ source terms. For observational comparison and testing its relevance, the Born--Infeld $F(T)$ solution models were applied on the Sgr$^*$A BH astrophysical system case by using the data-fitting techniques with light-deflection effects in ref. \cite{borninfeldadd}. In addition, the Born--Infeld models have also been tested on the disk accretion physical process implying BH systems in ref. \cite{baha6}. Ever if our new teleparallel $F(T)$ solutions are not exactly under the Born--Infeld form, we can claim that they are similar or comparable to the Born--Infeld BH solutions.

	Another possible physical process class developed in the recent literature are the scalarized BH solutions \cite{baha4,baha10,scalarizedbhnew}. This class of BH system is typically described by BHs evolving and in some case interacting inside an ambient scalar field such as a DE scalar field (quintessence, phantom or quintom). Several of the previous teleparallel $F(T)$ solutions may be relevant for this type of BH physical system. This is also another scope for the new fresh $F(T)$ solutions, especially for lower values of $b$-parameters. However, the scalarized BH models are usually more relevant for scalar-tensor ($F(T,\phi)$-gravity) and Gauss--Bonnet theories. But we can also treat this type of case with the new $F(T)$ solution as simple scalarized BH teleparallel $F(T)$ solutions. Finally, the observational relevance of the scalarized BH solutions is confirmed with the results of Sgr$^*$A and M87 BH shadow measurements in ref. \cite{scalarizedbhnew}. This is an additional proof of scalarized BH solutions pertinence in teleparallel~gravity.

	Technically, we used the Equation \eqref{powerlaw} coframe ansatz to find the most simple analytical new $F(T)$ solutions, but there are other possible ansatz approaches.} A good example is the partially flat exponential ansatz defined as $A_1(r)=a_0\left(1-e^{-ar}\right)$, {$A_2(r)=b_0=1$ and $A_3=c_0$} treated in detail in ref. \cite{nonvacSSpaper}. We set this ansatz so as to obtain a closed form which is leading to a Teleparallel Minkowski spacetime under the $r\,\rightarrow\,\infty$ limit in the definition of ref. \cite{landryvandenhoogen1}. The closed form coframe ansatz requirement is essential to really obtain new pure analytical teleparallel $F(T)$ solutions. From the last considerations, we studied in detail, from the new partially flat exponential ansatz $F(T)$ solutions, the possible singularities allowing for some new point-like astrophysical spacetime discontinuities. We had suggested that some hidden possible physical processes around the $F(T)$ function singularities can be studied in some future research works. This last aim is beyond the current paper's scope, but we can directly use the general Equation \eqref{2203} by substituting Equations \eqref{2922a}--\eqref{2922d} with the partially flat exponential ansatz for $A_1$--$A_3$ components to compute the new $F(T)$ solutions for any scalar field source described by a relevant $V(T)$ potential. Such a study may deserve an independent investigation as a good astrophysical application of static spherically symmetric teleparallel spacetime solution.

The last example of non-power-law coframe ansatz solution is not the only possible one. There are a great number of possible ansatzes other than Equation \eqref{powerlaw} and the partially flat exponential. And there are also other possible scalar field source definitions other than Equation \eqref{3000}. The next section will clarify this last point.

\section{Other Scalar Field Source Solutions}\label{sect4}

The Section \ref{sect3} aim was to find the new teleparallel $F(T)$ solutions for the power-law scalar field defined by Equation \eqref{3000} and the conservation law solutions of Equations \eqref{3004}, \eqref{3005} and \eqref{3006} for scalar potential $V(T)$. Another conclusion concerning the results obtained in Section \ref{sect3} is that the new $F(T)$ solutions are not only valid for a power-law-based scalar field, but also valuable for other scalar field definition and any types of potential $V(T)$ solutions of Equation \eqref{2201}. From this point, we are also able to use the same Equation \eqref{2203} for any type of scalar field source and potential by using the same coframe ansatz and the same subcases. We can keep the same teleparallel $F(T)$ solution forms, but we will change the scalar field definition, solve Equation \eqref{2201} and use the solutions in {Section \ref{sect3}} $F(T)$ by only changing the potential $V(T)$ expressions. This is the aim of this current section: we replace Equations \eqref{3004}, \eqref{3005} and \eqref{3006} by different form of scalar field $\phi(T)$ and potential $V(T)$ expressions. We will proceed under this approach for exponential and logarithmic scalar field definitions as relevant cases by the same manner as in recent papers for teleparallel cosmological spacetimes \cite{scalarfieldKS,scalarfieldTRW}.

\subsection{Exponential Scalar Field Solutions}\label{sect41}

The exponential scalar field is defined by
\begin{align}\label{4000}
	\phi(T)=p_0\,\exp\left(p\,r(T)\right){\equiv p_0\,\sum_{k=0}^{\infty}\,\frac{\left(p\,r(T)\right)^k}{k!}},
\end{align}
{where $k$ is a positive integer, $p$ and $p_0$ are {$\mathbb{R}$-valued} numbers. The exponential scalar field case can be considered as an infinite sum of power-law terms for completing {refs.~\cite{baha4,sharifmodels1,borninfeldadd}} models study in teleparallel gravity. As seen in Equation \eqref{borninfeldFT}, an exponential $F(T)$ from Born--Infeld BH class of solutions can be recovered in principle with a $r(T)\,\sim T$ relationship \cite{baha6,borninfeldadd}. The exponential scalar field defined by Equation \eqref{4000} will also be useful for scalarized BH solutions \cite{baha4,baha10,scalarizedbhnew}.} By assuming a power-law coframe ansatz, we will solve Equation \eqref{2201} to find the scalar potential $V(\phi)$ for {Equations \eqref{powerlaw} and \eqref{4000}}:
\begin{align}
	&\frac{dV}{d\phi}=p^2\,\phi+a\,p^2\,\frac{\phi}{\ln\left(\phi/p_0\right)},
	\nonumber\\
	&\Rightarrow\,V(\phi)=V_0+\frac{p^2}{2}\,\phi^{2}+a\,p^2\,Ei\left(2\ln\,\left(\phi/p_0\right)\right) . \label{4001}
\end{align}
{By} substituting Equation \eqref{4000} into Equation \eqref{4001} and setting $r=r(T)$, we obtain that
\begin{align}
	V(T)=V_0+\frac{p^2\,p_0^2}{2}\,\exp\left(2p\,r(T)\right)+a\,p^2\,Ei\left(2p\,r(T)\right) . \label{4002}
\end{align}
{{The} Equation \eqref{Quintessenceindex} for dark energy index is exactly:
	\begin{align}\label{4002a}
		\alpha_Q =& -1+\frac{p^2\,p_0^2\,\exp\left(2p\,r(T)\right)}{p^2\,p_0^2\,\exp\left(2p\,r(T)\right)+a\,p^2\,Ei\left(2p\,r(T)\right)+V_0}.
	\end{align}
	{Under} $p\,\rightarrow\,\infty$, Equation \eqref{4002a} will simplify as
	\begin{align}\label{4002b}
		\alpha_Q \approx & -1+\frac{p_0^2\,\exp\left(2p\,r(T)\right)}{p_0^2\,\exp\left(2p\,r(T)\right)+a\,Ei\left(2p\,r(T)\right)} \approx -1+1 \rightarrow \, 0 ,
	\end{align}
	where $Ei\left(2p\,r(T)\right)\ll \exp\left(2p\,r(T)\right)$ under the $p\,\rightarrow\,\infty$ limit. As for the power-law scalar field defined by Equation \eqref{3000}, we again obtain the \textbf{{ordinary matter limit}%MDPI: Please confirm if the bold is unnecessary and can be removed. AL: OK for bold format.
	}, because $P_{\phi}=0$ for any $\rho_{\phi}$ expression. For the DE processes, we need to satisfy the following constraints:
	\begin{enumerate}
		\item \textbf{{Quintessence:}} We need to require that $-1<\alpha_Q<-\frac{1}{3}$ from Equation \eqref{4002a}:
		\begin{align}
			p^2\,\exp\left(2p\,r(T)\right)- \frac{2a\,p^2}{p_0^2}\,Ei\left(2p\,r(T)\right) < & \frac{2V_0}{p_0^2},  \quad &  \alpha_Q<-\frac{1}{3}\; \text{requirement}, \label{4002c}
			\\
			-p^2\,\exp\left(2p\,r(T)\right)-\frac{a\,p^2}{p_0^2}\,Ei\left(2p\,r(T)\right) < & \frac{V_0}{p_0^2} \quad & \alpha_Q > -1\; \text{requirement}. \label{4002d}
		\end{align}
		{By} combining Equations \eqref{4002c} and \eqref{4002d}, we find in general that $a<0$ for positive $Ei\left(2p\,r(T)\right)$ and $V_0$ terms. The $a=-1$ and $-\frac{1}{2}$ classes of $F(T)$ solutions found in Sections \ref{sect323} and \ref{sect324} as the Sections \ref{sect321} and \ref{sect322} are in principle all ideal candidates for describing quintessence processes.
		
		\item \textbf{{Phantom Energy:}} We only require that $\alpha_Q<-1$ from Equation \eqref{4002a} as
		\begin{align}\label{4002e}
			p^2\,\exp\left(2p\,r(T)\right)- \frac{2a\,p^2}{p_0^2}\,Ei\left(2p\,r(T)\right)	<-\frac{V_0}{p_0^2}.
		\end{align}
		{In} this case, we have primarily that the $\frac{2a\,p^2}{p_0^2}\,Ei\left(2p\,r(T)\right)>p^2\,\exp\left(2p\,r(T)\right)$ and ideally that $-\frac{2a\,p^2}{p_0^2}\,Ei\left(2p\,r(T)\right)>p^2\,\exp\left(2p\,r(T)\right)$ term will be dominating for $a>0$. Only the general solution obtained in Section \ref{sect322} for positive values of $a$ can be a candidate for phantom energy models with an exponential scalar field.
	\end{enumerate}
	
	{In} %MDPI: We added indent here, please confirm. AL: OK for indent.
	Equations \eqref{4002c}--\eqref{4002e}, we have found the constraint for characteristic equation solution $r(T)$ and $p$ similar to Equations \eqref{3007a}--\eqref{3008c}.
	
}
\begin{enumerate}
	
	\item \textbf{{Power-law ansatz with} $A_3=c_0$}: We find the same $F(T)$ solution form as Equation~\eqref{3011} in Section \ref{sect31}, but the potential $V(T)$ will be Equation \eqref{4002} with $r(T)$ defined by Equation \eqref{3010}. Equation \eqref{4002} becomes:
	\begin{align}
		V(T)=V_0+\frac{p^2\,p_0^2}{2}\,\exp\left(\frac{2p\,\left(-\frac{2\delta\,c_0}{b_0}\right)^{\frac{1}{(1+b)}}}{{\left(u(T)\right)}^{\frac{1}{(1+b)}}}\right)+a\,p^2\,Ei\left(\frac{2p\,\left(-\frac{2\delta\,c_0}{b_0}\right)^{\frac{1}{(1+b)}}}{{\left(u(T)\right)}^{\frac{1}{(1+b)}}}\right) , \label{4003}
	\end{align}
	\normalsize
	{where $u(T)\neq 0$, $b_0\neq 0$, $c_0\neq 0$ and $b\neq -1$.}
	%r(T)=\left(-\frac{\delta\,b_0}{2c_0}\right)^{-\frac{1}{(1+b)}}{\left(u(T)\right)}^{-\frac{1}{(1+b)}}
	
	\item \textbf{{Power-law ansatz with} $A_3=r$}: We find the same $F(T)$ solution forms as in Section~\ref{sect32}, but only the $V(T)$ potential expressions change by replacing {Equations \eqref{3004}, \eqref{3005} and \eqref{3006}} by Equation \eqref{4002} for each subcase treated in this section. Equation \eqref{4002} will be for the simplest cases:
	\begin{enumerate}
		\item {${\bf a=b=0}$}: The potential $V(T)$ in Equation \eqref{3104} is
		\begin{align}
			V(T)=V_0+\frac{p^2\,p_0^2}{2}\,\exp\left(\frac{2\sqrt{2}\,p\left(1+\delta\,b_0\right)}{b_0\,(-T)^{1/2}}\right) , \label{4011}
		\end{align}
		{where $T<0$ and $b_0\neq \left\lbrace 0,\,-\delta \right\rbrace$.}
		
		\item {${\bf a\neq \left\lbrace -1,\, -\frac{1}{2} \right\rbrace}$} and {${\bf b=0}$}: The $V(T)$ expression in Equation \eqref{3114} is
		\begin{align}
			V(T)=& V_0+\frac{p^2\,p_0^2}{2}\,\exp\left(\frac{2\sqrt{2}\,p\left(b_0^2+2\delta\,b_0\,(1+a)+2\,a+1\right)^{1/2}}{b_0\left(-T\right)^{1/2}}\right)
			\nonumber\\
			&\;+a\,p^2\,Ei\left(\frac{2\sqrt{2}\,p\left(b_0^2+2\delta\,b_0\,(1+a)+2\,a+1\right)^{1/2}}{b_0\left(-T\right)^{1/2}}\right) , \label{4012}
		\end{align}
		{where $T<0$ and $b_0\neq \left\lbrace 0,\,-\delta \right\rbrace$.}
		
		\item {${\bf a\neq \left\lbrace -1,\, -\frac{1}{2} \right\rbrace}$} and {${\bf b=-1}$}: The $V(T)$ expression in Equation \eqref{3134} {is described by Equation \eqref{4002} with $r(T)$ defined by Equation \eqref{3132}. The other cases will use Equation \eqref{4002} with the appropriate $r(T)$ expression depending on the value of $b$.}
		
		\item {${\bf a=-1}$} and {${\bf b=0}$}: The $V(T)$ expression in Equation \eqref{3154} is
		\vspace{-10pt}
%		\begin{adjustwidth}{-\extralength}{0cm}
			%\centering %% If there is a figure in wide page, please release command \centering
			\begin{align}
				V(T)=V_0+\frac{p^2\,p_0^2}{2}\,\exp\left(\frac{2\sqrt{2}\,p\left(1-b_0^2\right)^{1/2}}{b_0\,T^{1/2}}\right)-p^2\,Ei\left(\frac{2\sqrt{2}\,p\left(1-b_0^2\right)^{1/2}}{b_0\,T^{1/2}}\right) , \label{4014}
			\end{align}
%		\end{adjustwidth}
		{where $T>0$ and $b_0\neq \left\lbrace 0,\,\delta \right\rbrace$.}
		
		\item {${\bf a=-1}$} and {${\bf b=-1}$}: The $V(T)$ expression in Equation \eqref{3158} is
		\vspace{-10pt}
%		\begin{adjustwidth}{-\extralength}{0cm}
			%\centering %% If there is a figure in wide page, please release command \centering
			\begin{align}
				V(T)=V_0+\frac{p^2\,p_0^2}{2}\,\exp\left(2p\,\left[\frac{1}{b_0^2}-\frac{T}{2}\right]^{-1/2}\right)-p^2\,Ei\left(2p\,\left[\frac{1}{b_0^2}-\frac{T}{2}\right]^{-1/2}\right) , \label{4015}
			\end{align}
%		\end{adjustwidth}
		{where $\frac{2}{b_0^2}>T$ and $b_0\neq 0$.}
		
		\item {${\bf a=-\frac{1}{2}}$} and {${\bf b=0}$}: The $V(T)$ expression in Equation \eqref{3243} is
		\vspace{-10pt}
%		\begin{adjustwidth}{-\extralength}{0cm}
			%\centering %% If there is a figure in wide page, please release command \centering
			\begin{align}
				V(T)=V_0+\frac{p^2\,p_0^2}{2}\,\exp\left(2p\,\left[-{\frac{2\left(b_0+\delta\right)}{b_0\,T}}\right]^{1/2}\right)-\frac{p^2}{2}\,Ei\left(2p\,\left[-{\frac{2\left(b_0+\delta\right)}{b_0\,T}}\right]^{1/2}\right) , \label{4016}
			\end{align}
%		\end{adjustwidth}
		{where $T<0$ and $b_0\neq \left\lbrace 0,\,-\delta \right\rbrace$.}
		
		\item {${\bf a=-\frac{1}{2}}$} and {${\bf b=-1}$}: The $V(T)$ expression in Equation \eqref{3247} is
		\begin{align}
			V(T)=& V_0+\frac{p^2\,p_0^2}{2}\,\exp\left(2p\,\left[-\frac{\delta}{2b_0}\pm\sqrt{\frac{1}{4b_0^2}-\frac{T}{2}}\right]^{-1}\right)
			\nonumber\\
			&\;-\frac{p^2}{2}\,Ei\left(2p\,\left[-\frac{\delta}{2b_0}\pm\sqrt{\frac{1}{4b_0^2}-\frac{T}{2}}\right]^{-1}\right) , \label{4017}
		\end{align}	
		{where $\frac{1}{2b_0^2}\geq T$ and $b_0\neq 0$.}
		
	\end{enumerate}
	
	{The} %MDPI: We added indent here, please confirm. AL: OK
	other subcases of Section \ref{sect32} can be computed by the same manner as in the previous simple examples. 
	
\end{enumerate}

\subsection{Logarithmic Scalar Field Solutions}\label{sect42}

{As performed in refs. \cite{baha4,sharifmodels1}, the logarithmic} scalar field is defined by
\begin{align}\label{5000}
	\phi(T)=p_0\,\ln\left(p\,r(T)\right).
\end{align}
{where $p$ and $p_0$ are $\mathbb{R}$-valued. The scalar field definition of Equation \eqref{5000} is similar to those used for scalarized BH solutions in ref. \cite{baha4} and constitutes an additional proof of Equation~\eqref{5000}'s relevance. By assuming a power-law coframe ansatz}, we will solve Equation \eqref{2201} to find the scalar potential $V(\phi)$ for Equations \eqref{powerlaw} and \eqref{5000}:
\begin{align}
	&\frac{dV}{d\phi}=(a-1)\,p_0\,p^2\,\exp\left(-2\phi/p_0\right),
	\nonumber\\
	&\Rightarrow\,V(\phi)= V_0+\frac{(1-a)}{2}\,p_0^2\,p^2\,\exp\left(-2\phi/p_0\right) , \label{5001}
\end{align}
{where $a\neq 1$.} By substituting Equation \eqref{5000} into Equation \eqref{5001} and setting $r=r(T)$, we obtain that
\begin{align}
	V(T)= V_0+\frac{(1-a)}{2}\,p_0^2\,r^{-2}(T) . \label{5002}
\end{align}
Equation \eqref{Quintessenceindex} for dark energy index becomes
\begin{align}\label{5002a}
	\alpha_Q =&  -1+\left[1-\frac{a}{2}+\frac{V_0}{p_0^2}\,r^2(T)\right]^{-1}.
\end{align}	
{For} $a=2$, Equation \eqref{5002a} simplifies to $\alpha_Q =-1+\frac{p_0^2}{V_0}\,r^{-2}(T)$ for any value of $p$.

{The scalar field definition of Equation \eqref{5000} also represents a good dilaton model as shown in ref. \cite{sharifmodels1}. Therefore, we also need to set some requirements for DE-based~solutions:}
\begin{enumerate}
	\item  \textbf{{Quintessence:}} We need that $-1<\alpha_Q<-\frac{1}{3}$ for this type of physical process and Equation \eqref{5002a} leads to
	\begin{align}
		\frac{(a+1)\,p_0^2}{2V_0}< & r^2(T), \quad  & \alpha_Q<-\frac{1}{3}\;\text{requirement},\label{5002b}
		\\
		\frac{(a-2)\,p_0^2}{2V_0}< & r^2(T)  , \quad  & \alpha_Q > -1\;\text{requirement}, \label{5002c}
	\end{align}	
	{By} satisfying the Equation \eqref{5002b} criterion, we automatically satisfy Equation \eqref{5002c}'s requirement. To guarantee a $\mathbb{R}$-valued $r(T)$ expression, it requires that $a> -1$ in {Equations \eqref{5002b} and \eqref{5002c}} for a quintessence process. For the $a=2$ subcase, {Equation~\eqref{5002b}}'s constraint simplifies as $\frac{3p_0^2}{2V_0}< r^2(T)$. Any section \ref{sect32} may lead in principle to quintessence~solution.
	
	\item \textbf{{Phantom Energy:}} Equation \eqref{5002a} with $\alpha_Q<-1$ requirement leads to the following constraint:
	\begin{align}
		r^2(T)	< 	\frac{(a-2)\,p_0^2}{2V_0}
	\end{align}
	{For} any positive $V_0$ and $r^2(T)$ solution (or $\mathbb{R}$-valued), we need to satisfy the $a>2$ criterion. In a such case, only the Section \ref{sect322} solutions may lead to the phantom energy models and any Sections \ref{sect323} and \ref{sect324} cannot lead to this type of models (due to the negative values of $a$).
\end{enumerate}

\begin{enumerate}
	
	\item \textbf{{Power-law ansatz with} $A_3=c_0$}: As in Section \ref{sect41}, the $F(T)$ solution is under the same form as Equation \eqref{3011} in Section \ref{sect31}. The potential $V(T)$ will be{Equation \eqref{5002}} with $r(T)$ defined by Equation~\eqref{3010} as
	\begin{align}
		V(T)= V_0+\frac{(1-a)}{2}\,p_0^2\,\left(\frac{b_0}{2c_0}\right)^{\frac{2}{(1+b)}}{\left(u(T)\right)}^{\frac{2}{(1+b)}} , \label{5003}
	\end{align}
	{where $u(T)\neq 0$, $b_0\neq 0$, $c_0\neq 0$ and $b\neq -1$.}
	%r^{-2}(T)=\left(-\frac{\delta\,b_0}{2c_0}\right)^{\frac{2}{(1+b)}}{\left(u(T)\right)}^{\frac{2}{(1+b)}}
	
	\item \textbf{{Power-law ansatz with} $A_3=r$}: As in Section \ref{sect41}, we find the same $F(T)$ solution forms than Section \ref{sect32}, but only the $V(T)$ potential expressions change for the {Equation \eqref{5002}} form for each subcases treated in this section. Equation \eqref{5002} will be for the simplest cases:
	\begin{enumerate}
		\item {${\bf a=b=0}$}: The potential $V(T)$ in Equation \eqref{3104} is
		\begin{align}
			V(T)= V_0+\frac{p_0^2\,b_0^2}{4\left(1+\delta\,b_0\right)^2} (-T) , \label{5011}
		\end{align}
		{where $T \neq 0$ and $b_0\neq \left\lbrace 0,\,-\delta \right\rbrace$.}
		
		\item {${\bf a\neq \left\lbrace -1,\, -\frac{1}{2} \right\rbrace}$} and {${\bf b=0}$}: The $V(T)$ expression in Equation \eqref{3114} is
		\begin{align}
			V(T)= V_0+\frac{p_0^2\,b_0^2\,(1-a)}{4\,\left(b_0^2+2\delta\,b_0\,(1+a)+2\,a+1\right)}\,\left(-T\right) , \label{5012}
		\end{align}
		{where $T \neq 0$ and $b_0\neq \left\lbrace 0,\,-\delta \right\rbrace$.}
		
		\item {${\bf a\neq \left\lbrace -1,\, -\frac{1}{2} \right\rbrace}$} and {${\bf b=-1}$}: The $V(T)$ expression in Equation \eqref{3134} {is described by Equation \eqref{5002} with $r(T)$ defined by Equation \eqref{3132}. The other cases use Equation \eqref{5002} with the $r(T)$ expression for the appropriate value of $b$.}
		
		\item {${\bf a=-1}$} and {${\bf b=0}$}: The $V(T)$ expression in Equation \eqref{3154} is
		\begin{align}
			V(T)= V_0+ \frac{p_0^2\,b_0^2}{2\left(1-b_0^2\right)}\,T , \label{5014}
		\end{align}
		{where $T \neq 0$ and $b_0\neq \left\lbrace 0,\, \delta\right\rbrace $.}
		
		\item {${\bf a=-1}$} and {${\bf b=-1}$}: The $V(T)$ expression in Equation \eqref{3158} is
		\begin{align}
			V(T)= V_0+p_0^2\,\left[\frac{1}{b_0^2}-\frac{T}{2}\right] , \label{5015}
		\end{align}
		{where $T \neq 0$ and $b_0\neq 0$.}
		
		\item {${\bf a=-\frac{1}{2}}$} and {${\bf b=0}$}: The $V(T)$ expression in Equation \eqref{3243} is
		\begin{align}
			V(T)= V_0-\frac{3p_0^2\,b_0}{8\left(b_0+\delta\right)}\,T , \label{5016}
		\end{align}
		{where $T \neq 0$ and $b_0\neq \left\lbrace 0,\, -\delta\right\rbrace $.}
		
		\item {${\bf a=-\frac{1}{2}}$} and {${\bf b=1}$}: The $V(T)$ expression in Equation \eqref{3247} is
		\begin{align}
			V(T)= V_0+\frac{3}{4}\,p_0^2\,\left[-\frac{\delta}{2b_0}\pm\sqrt{\frac{1}{4b_0^2}-\frac{T}{2}}\right]^2 , \label{5017}
		\end{align}
		{where $\frac{1}{2b_0^2}\geq T$ and $b_0\neq 0$.}
		
	\end{enumerate}
	
	{As} %MDPI: We added indent here, please confirm. AL: OK for indent.
	in Section \ref{sect41}, the other subcases of Section \ref{sect32} can be computed in the same manner as in the current section examples.
	
\end{enumerate}

{
	\subsection{Graphical Comparisons and Summary of Main Results}\label{sect43}

	In addition to finding a large number of new analytic teleparallel $F(T)$ solutions, it is relevant to compare some of these simpler solutions as a final step of this development work by using some plots. This will make the inter-solution comparisons easier. We will restrict the comparison to Equation \eqref{powerlaw} power-law ansatz with $A_3=r$ for $a=b=0$, $b=0$ general and $b=-1$ subcases and use Equations \eqref{3004}, \eqref{3006}, \eqref{4002} and \eqref{5002} for, respectively, power-law general, ordinary matter  $p\gg 1$ limit, exponential and logarithmic scalar field potential.

	For comparing the $a=b=0$ case solutions, the $F(T)$ solution described by {Equation~\eqref{3104}} with potential $V(T)$ is defined by Equations \eqref{3105}, \eqref{3106}, \eqref{4011} and \eqref{5011}. The plots are presented in Figure \ref{figure1} and compare the different cases depending on the potential $V(T)$ expression. This is the most simple case, because we have essentially a two power-law terms superposition as $F(T)$ solution. The $p\gg1$ limit curve shows that there is a maximal value of $T=T_0\approx -5$ for this case but for $T< -5$, the $F(T)$ solution curve follows the negative exponential curve and a the power-law plot slightly for lower values of $T$. The logarithmic curve will take a gradual shift compared to the ordinary matter limit, power-law and negative exponential curves, but follows a similar type of curve. The positive exponential curve is an extreme limit of the power-law curve, because this can be considered as an infinite superposition of power-law terms. 
	\begin{figure}[h]%
		\includegraphics[scale=0.8]{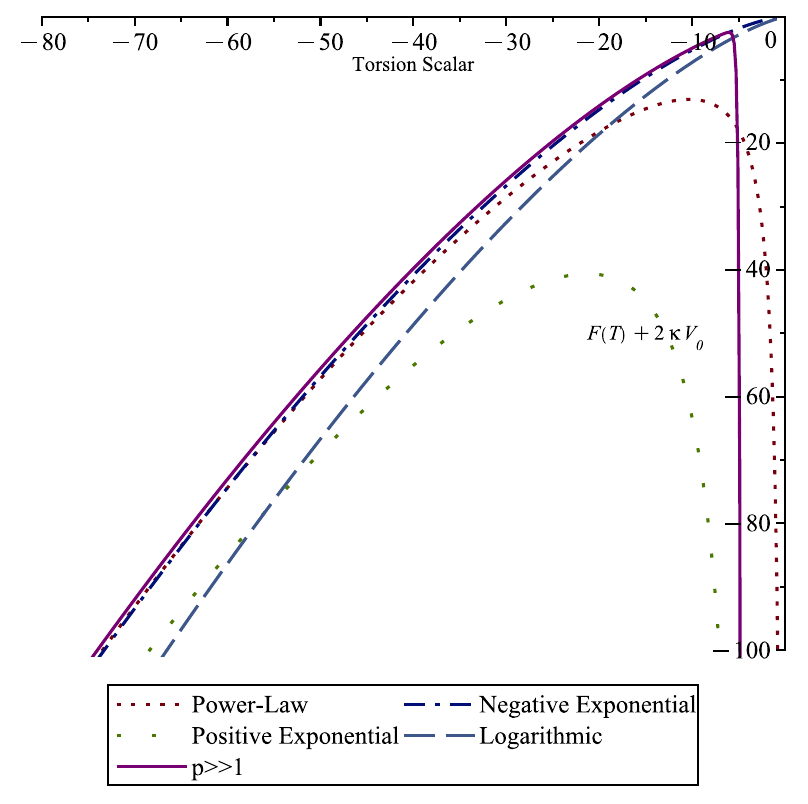}
		\caption{Plot of $F(T)$ versus $T$ for pure flat cosmological $a=b=0$ subcase ($p_0^2=1$, $b_0=2\delta$, $p=2$ (except for $p\gg 1$ case)).}
		\label{figure1}
	\end{figure}

	For the general $b=0$ (and $a\neq 0$) case, Figure \ref{figure2} shows that $F(T)$ solutions are described by Equation \eqref{3114} with $V(T)$ potential defined by {Equations \eqref{3115}, \eqref{3117}, \eqref{4012} and \eqref{5012}} for power-law, ordinary matter limit, exponential and logarithmic $V(T)$ potentials. Once again, we can see for the ordinary matter limit case a maximal value of $T=T_0\approx -20$, and for $T<-20$ the curve follows about the same path as the negative exponential as in Figure \ref{figure1}. The power-law curve has the same form than that in Figure \ref{figure1}, only the shifting and scale are a bit different between the Figures \ref{figure1} and \ref{figure2} cases. The positive exponential case does not appear on the figure, because the curve is too far from the other ones. Finally, the logarithmic curve is ever shifted from the other curves, but to the inverse side compared to Figure \ref{figure1}. Figure \ref{figure2} curves can be seen as a generalization of Figure \ref{figure1} for the non-zero $a$-parameter.
	\begin{figure}[h]%
		\includegraphics[scale=0.8]{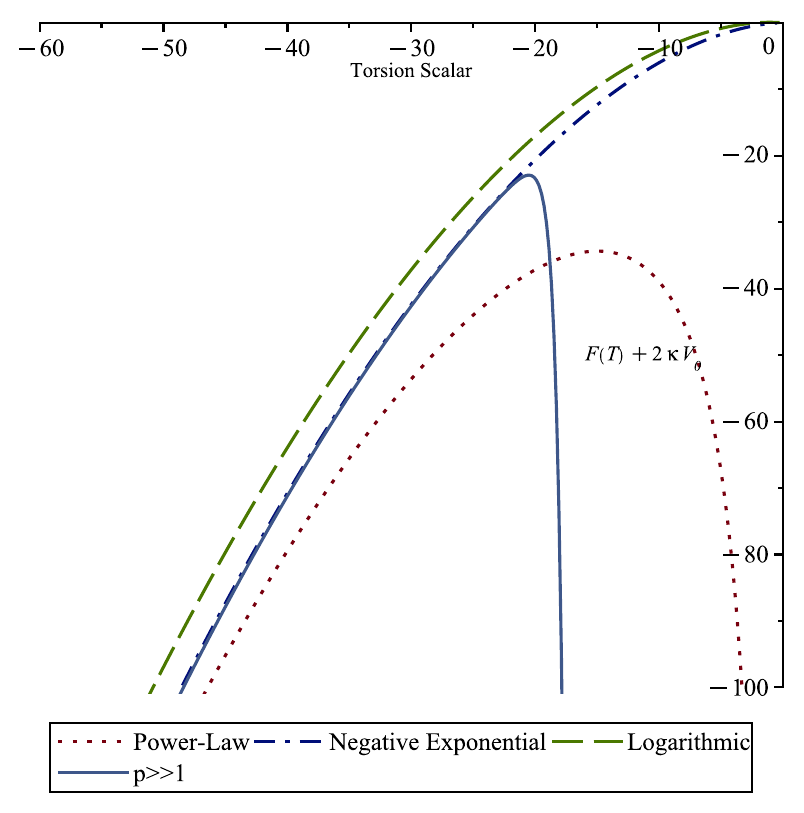}
		\caption{Plot of $F(T)$ versus $T$ for $b=0$ and $a\neq 0$ subcase ($p_0^2=1$, $b_0=2\delta$, $p=2$ (except for $p\gg 1$~case)).}
		\label{figure2}
	\end{figure}
	
	Figure \ref{figure3} essentially shows that all curves are equivalent at the large scale for the $b=-1$ and $a\neq 0$ subcase and for any potential case.%EE: Please ensure that the intended meaning has been retained. AL: OK.
	The main reason is that the geometry terms (non-$V(T)$ terms) are widely dominating compared to the potential $V(T)$ source term. This situation is also occurring for several higher values of $b$-parameter cases found in this work, but also in the recent studies of Born--Infeld $F(T)$ solution models in refs. \cite{baha6,borninfeldadd}. This $b=-1$ case could also picture the higher values of $b$ teleparallel $F(T)$ solutions. To make the distinction between potential cases, we plotted on Figure \ref{figure3a} the $V(T)$ function to highlight the difference between curves. We can see for the ordinary matter limit that the maximal value of $r(T)$ is $r(T)= 0.93$ in Figure \ref{figure3a}. We can also see that the power-law and logarithmic curves are increasing, but the exponential case curve increases much more in comparison with the other curves. The logarithmic and exponential plots are asymptotic in comparison to the power-law one. All the $b=-1$ subcase curve description can also represent those for higher values of $b$ subcases, because the potential terms can be considered as small corrections in such typical $F(T)$ solutions.
	\begin{figure}[h]%
		\includegraphics[scale=0.85]{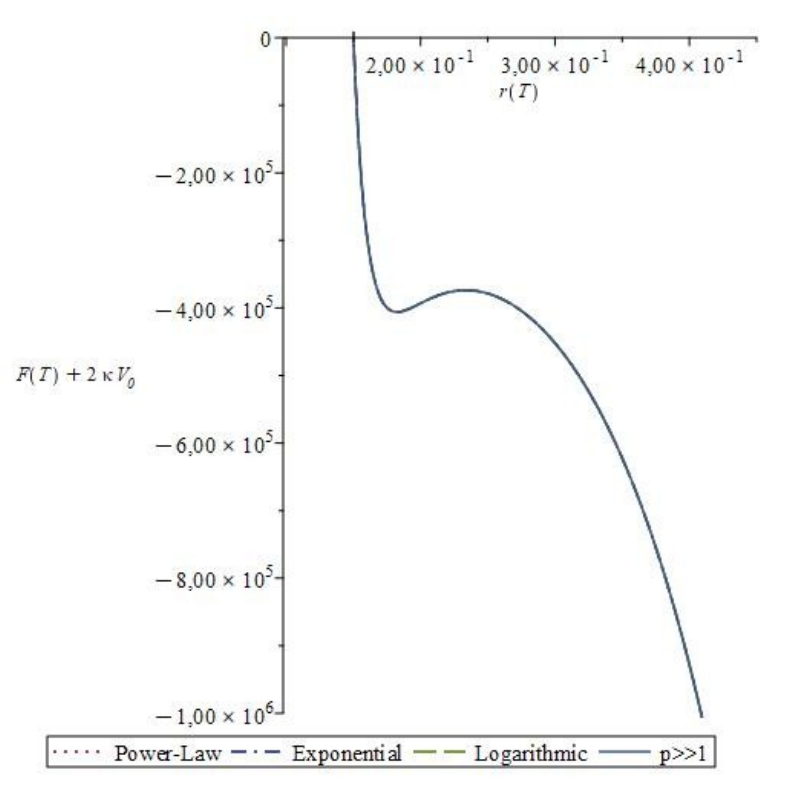}
		\caption{{Plot} of $F(T)$ versus $r(T)$ for $b=-1$ and $a\neq 0$ subcase ($p_0^2=1$, $b_0=2\delta$, $p=2$ (except for $p\gg 1$ case)). All other $b\neq 0$ will be similar to this last case.}
		\label{figure3}
	\end{figure}
	\vspace{-8pt}
	\begin{figure}[h]%
		\includegraphics[scale=0.8]{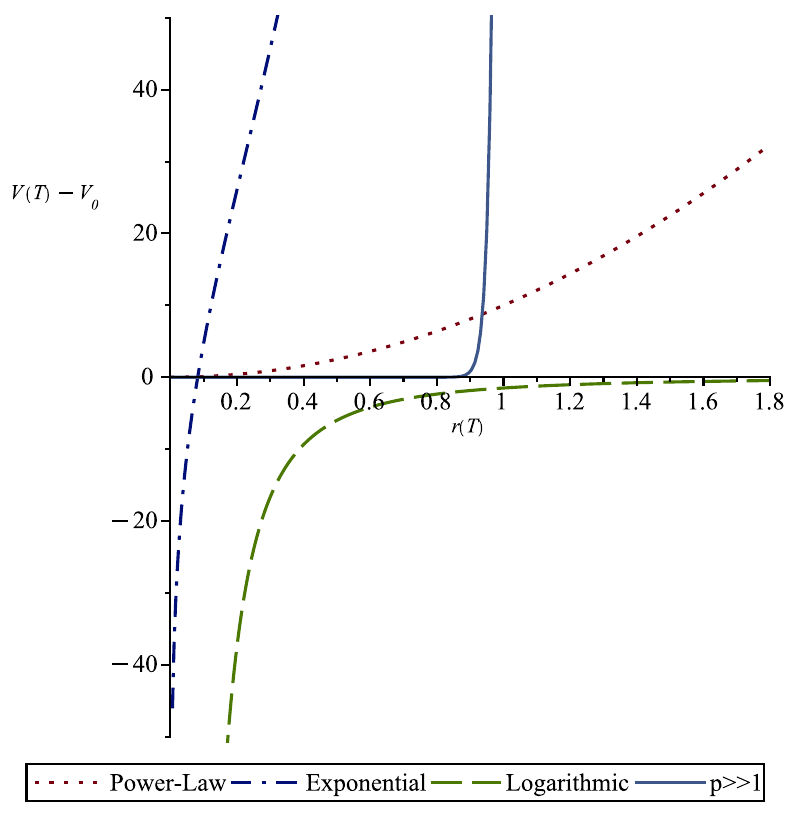}
		\caption{Plot of $V(T)$ versus $r(T)$ for $b=-1$ and $a\neq 0$ subcase ($p_0^2=1$, $b_0=2\delta$, $p=2$ (except for $p\gg 1$ case)).}
		\label{figure3a}
	\end{figure}

	We could perform with additional graph plots, but the analysis steps and results will be similar to those found in Figures \ref{figure1}--\ref{figure3a}. The higher values of $b$ subcases will be similar to the plots in Figures \ref{figure3} and \ref{figure3a} and the lower values of $b$ will look like Figures \ref{figure1} and \ref{figure2}, without major differences.

	\subsubsection*{Summary of Main Analytical Solutions}
	
	\begin{enumerate}

		\item $A_3=c_0$ : The main teleparallel $F(T)$ solution is described by Equation \eqref{3011} with potentials $V(T)$ defined by Equations \eqref{3013}--\eqref{3017} for any values of $a$ and $b$.

		\item $A_3=r$ The teleparallel $F(T)$ solutions cases are summarized in Table \ref{table1} in terms of $a$ and $b$.

		\begin{table}[ht]
			\begin{tabular}{ccccc}
				\hline
				$a$ &\hspace*{0.5cm} $0$ &\hspace*{0.5cm} $\neq 0$ &\hspace*{0.5cm} $-1$ & \hspace*{0.5cm} $-\frac{1}{2}$ \\
				\hline
				$b$ &\hspace*{0.5cm} $0$ &\hspace*{0.5cm} $0$, $\pm 1$, $-2$ &\hspace*{0.5cm} $0$, $\pm \frac{1}{2}$, $\pm 1$, $-\frac{3}{2}$ $\pm 2$, $3$   & \hspace*{0.5cm} $0$, $\pm 1$, $\pm 2$, $-3$, $\pm 4$, $\pm 6$, $-8$\\	
				\hline
			\end{tabular}
			\caption{{Summary} of the $b$-parameter values with a $A_3=r$ power-law coframe ansatz leading to analytical teleparallel $F(T)$ solutions for any scalar field potential $V(T)$.}
			\label{table1}
		\end{table}
		
	\end{enumerate}
	
}
\newpage

\section{Concluding Remarks}\label{sect5}

We have treated, simplified and solved the static teleparallel $F(T)$ FEs by summarizing everything to the single Equation \eqref{2203} to obtain the teleparallel $F(T)$ solutions. We have obtained solutions for subcases similar to those treated in ref. \cite{nonvacSSpaper} by replacing the astrophysical perfect fluid by a scalar field as energy-momentum source with only the radial dependence ($r$-coordinate). This scalar field source and then the conservation law solutions (described by Equation \eqref{2201} for the potential $V(T)$) can represent a local source of DE where an astrophysical system evolves. We obtain teleparallel $F(T)$ solutions with several common points with those of the linear perfect fluids found in ref. \cite{nonvacSSpaper}. More concretely, Equation \eqref{2203} makes it possible to generate all $F(T)$ solutions for any type of ansatz, scalar field $\phi(T)$ and any type of scalar potential $V(T)$ solutions of Equation \eqref{2201}. By this approach, the general forms of the $F(T)$ solutions obtained in Section \ref{sect3} remain intact, regardless of the potential $V(T)$. This constitutes a major strength favoring the approach based on {Equation~\eqref{2203}}. Section \ref{sect4} shows that one can also replace the power-law scalar field $\phi(T)$ used in Section \ref{sect3} by exponential and logarithmic $\phi(T)$ (Sections \ref{sect41} and \ref{sect42}, respectively) while keeping the same forms of $F(T)$ solutions as in Section \ref{sect3} and respecting Equation~\eqref{2201}. {However, we also find the specific requirements for  quintessence and phantom DE forms to satisfy for the $r(T)$ characteristic equation solutions and the $p$-parameters of each type of scalar field source definitions. By this manner, we can claim that any new teleparallel $F(T)$ solution found here is in principle applicable for quintessence models, but only the Section \ref{sect322} new $F(T)$ solutions can be applicable for phantom DE models.} {The {Sections \ref{sect323} and \ref{sect324}} cases ($a=-1$ and $-\frac{1}{2}$, respectively) allow for a large number of easy-to-compute new teleparallel $F(T)$ solutions describing the astrophysical systems evolving in a DE quintessence scalar field.} In short, this work simultaneously and logically completes the advances made in ref. \cite{nonvacSSpaper}.

All {these new developments} will make the use of teleparallel $F(T)$ solutions and scalar field sources easier {for future astrophysical applications involving systems immersed and evolving in a DE-dominating universe that can fundamentally be described by a scalar field. There are some recent papers addressing two types of astrophysical applications in teleparallel gravity using some similar approaches to the current paper: the Born--Infeld BH (see refs. \cite{baha6,borninfeldadd}) and the scalarized BH solutions (see refs. \cite{baha4,baha10,scalarizedbhnew}). We see that the Born--Infeld models may be more suitable to the higher values of $b$ cases and the scalarized BH models may be suitable for lower values of $b$ cases. In addition, these two same models were tested and compared with real astrophysical situations with observation data confirming the rightness of these models. Under all the previous considerations, the Born--Infeld and scalarized BH models and solutions can be used as suitable comparison points for future detailed works on specific teleparallel $F(T)$ BH solutions.}

We should note that a relatively similar approach had been used in several recent works in teleparallel $F(T)$ gravity, {particularly} for KS and TRW spacetimes depending only on the time coordinate \cite{scalarfieldKS,scalarfieldTRW}. These latest advances have at the same time laid the foundations for this present paper by clarifying some points on the DE. However, the present contribution with its new teleparallel $F(T)$ solutions will make it possible to deal much more easily and realistically with the various types of possible astrophysical systems in the DE, harmonizing better with cosmological models. This has become possible with the powerful tools of teleparallel gravity.

\section*{Abbreviations}

\noindent The following abbreviations are used in this manuscript:\\
\noindent 
\begin{tabular}{@{}ll}
	AL & Alexandre Landry \\	
DE & Dark Energy  \\
EoS & Equation of State 	  \\
GR & General Relativity 	 \\
KV & Killing Vector 	 \\
NS & Neutron star  \\
TRW & Teleparallel Robertson--Walker  \\
BH & Black Hole\\
DoF & Degree of Freedom\\
FE & Field Equation\\
KS & Kantowski--Sachs\\
NGR & New General Relativity\\
TEGR & Teleparallel Equivalent of General Relativity\\
WD & White Dwarf \\
\end{tabular}

%\vspace*{0.5cm}
\newpage

\section*{Notation}

\begin{tabular}{@{}ll}
$\mu, \nu, \ldots$ & coordinate indices \\	
$a,b,\ldots$ & 	tangent space indices \\
$x^\mu$ & spacetime coordinates \\
${\bf h}_a$,  ${\bf h}^a$, $h_a^{~~\mu}$ or $h^a_{~~\mu}$  & coframe expressions (tetrad for orthonormal frames) \\
$\omega^a_{~~bc}$, $\omega^a_{~~b\mu}=\omega^a_{~~bc}h^c_{~\mu}$	& spin-connection \\
$g_{ab}$ & gauge metric \\
$g_{\mu \nu}=g_{ab}\,h^a_{~\mu}h^b_{~\nu} $ & spacetime metric \\
$R^a_{~bcd}$ & curvature tensor \\
$T^a_{~bc}$ & torsion tensor \\
$T$ & torsion scalar \\
$S_a^{~~\mu\nu}$ & superpotential \\
$F=F(T)$ & teleparallel theory function of $T$ \\
$F_T=\frac{dF(T)}{dT}$, $F_{TT}=\frac{d^2F(T)}{dT^2}$ & derivatives with respect to (w.r.t) $T$ \\
$\overset{\ \circ}{G}_{ab}$ & Einstein tensor \\
$\Theta_a^{~~\mu}$, $\Theta^{\mu\nu}$ & conserved energy-momentum tensor \\
$\overset{\ \circ}{\nabla}\,_{\nu}$ & covariant derivative \\
$\Lambda^a_{~b}$ & Lorentz Transformation \\
$\mathfrak{T}_{ab}$ & hypermomentum  \\
$\phi$ & scalar field \\
$A'=\partial_r\,A$ & radial coordinate derivative \\

\end{tabular}

%\vspace*{0.5cm}
%\newpage

%\newpage

\appendix

\section{Field Equation Components}\label{appena}
%\subsection[\appendixname~\thesubsection]{}

{There are, in this appendix, the necessary FE components for general, constant $A_3$ and $A_3=r$ power-law coframe ansatz cases. These useful components for the paper's purposes have been found and used in refs. \cite{SSpaper,nonvacSSpaper}. }

\subsection{General Components}
%\vspace{-10pt}
%\small
%\begin{adjustwidth}{-\extralength}{0cm}
%\centering %% If there is a figure in wide page, please release command \centering
\begin{align}
	\frac{g_1}{k_1}=&\frac{\Bigg[-A_2\,A_3^2\,A_1''-A_1\,A_2\,A_3\,A_3''+A_1\,A_2\,A_3'^2+\left(A_1\,A_2\right)'\,A_3\,A_3'+A_3^2\,A_1'\,A_2'-A_1\,A_2^3\Bigg]}{\left[A_1\,A_2\,A_3\,A_3'+A_2\,A_3^2\,A_1'+\delta\,A_1\,A_2^2\,A_3\right]} \label{2922a}
	\\
	g_2 =& \frac{1}{A_1\,A_2^3\,A_3}\Bigg[-A_1\,A_2\,A_3''+\left(A_1\,A_2\right)'\,A_3'\Bigg] \label{2922b}
	\\
	g_3 =& \frac{1}{A_1\,A_2^3\,A_3^2}\Bigg[-A_1\,A_2\,A_3\,A_3''-A_1\,A_2\,A_3'^2-A_2\,A_3\,A_3'\,A_1'-\delta\,A_1\,A_2^2\,A_3'+A_1\,A_3\,A_2'\,A_3'
	\nonumber\\
	&\quad-\delta\,A_2^2\,A_3\,A_1'\Bigg] ,\label{2922c}
	\\
	k_2=& \frac{1}{A_2^2\,A_3}\left[A_3'+\delta\,A_2\right] . \label{2922d}
\end{align}
%\end{adjustwidth}
%\normalsize

\subsection{$A_3=c_0=$ Constant Power-Law Components}

The Equations \eqref{2922a}--\eqref{2922d} with the Equations \eqref{powerlaw} power-law ansatz are
%\small
%\vspace{-10pt}
%\begin{adjustwidth}{-\extralength}{0cm}
%\centering %% If there is a figure in wide page, please release command \centering
\begin{align}\label{2923}
	&\frac{g_1}{k_1} = \frac{\left[\left(a(1-a+b)\right)\,r^{-2b-2}-\left(\frac{b_0}{c_0}\right)^2\right]}{\left[a\,r^{-b-1}+\left(\frac{\delta\,b_0}{c_0}\right)\right]\,r^{-b}} ,\quad  g_2=0 , \quad g_3 = -\left(\frac{\delta\,a}{b_0\,c_0}\right)\,r^{-b-1} , 
	\quad k_2=\left(\frac{\delta}{b_0\,c_0}\right)\,r^{-b}  .
\end{align}
%\end{adjustwidth}
%\normalsize

\subsection{$A_3=r$ Power-Law Components}

The Equations \eqref{2922a}--\eqref{2922d} with Equations \eqref{powerlaw} power-law ansatz are
%\small
%\vspace{-10pt}
%\begin{adjustwidth}{-\extralength}{0cm}
%\centering %% If there is a figure in wide page, please release command \centering
\begin{align}\label{2924}
	&\frac{g_1}{k_1} = \frac{\left[{B_{ab}}\,r^{-2b}-b_0^2\right]}{\left[(a+1)\,r^{2(1-b)-1}+\delta\,b_0\,r^{(1-b)}\right]} ,~
	& g_2 =& \frac{(a+b)}{b_0^2}\,r^{-2b-2} , &
	\nonumber\\
	& g_3 = \frac{1}{b_0^2}\Bigg[(-a+b-1)\,r^{-2b-2}-\delta\,b_0\,(a+1)\,r^{-b-2}\Bigg] ,~ & k_2=& \frac{1}{b_0^2}\Bigg[r^{-2b-1}+\delta\,b_0\,r^{-b-1}\Bigg] . &
\end{align}
%\end{adjustwidth}
%\normalsize

%\newpage

{
	\section{Tables of Sections \ref{sect323} and \ref{sect324} Special Functions and~Solutions}\label{appenb}
	
	This appendix presents the computed values of Equations \eqref{3158function} and \eqref{3247function}, supplemented by the necessary Equation \eqref{3100} $r(T)$ solutions for the computations of each subcase.
	
	\begin{table}[h]%[ht]
		\caption{{The} Equation \eqref{3100} $r(T)$ solutions where $a=-1$. Note that $\left(\delta_1,\,\delta_2\right)=\left(\pm 1,\,\pm 1\right)$ for any positive $r(T)$.}
		\label{tableb1}
		\begin{tabular}{ccc}
			\hline
			\boldmath{$b$}& \boldmath{$r(T)$} & \boldmath{$x_b(T)$} \\
			\hline
			$\frac{1}{2}$	& $3 \Bigg[b_0^2+\frac{2^{2/3}\,b_0^4}{\sqrt[3]{x_{1/2}(T)}}+\frac{\sqrt[3]{x_{1/2}(T)}}{2^{2/3}}\Bigg]^{-1}$ & $3^{3/2}\sqrt{27b_0^4\,T^2+8b_0^8\,T}+27b_0^2\,T+4b_0^6$ \\
			\hline
			$-\frac{1}{2}$	& ${2b_0^2}\left[1 +\delta_1 \sqrt{1-2\,b_0^4\,T}\right]^{-1}$ & {N.A.} %MDPI: Please check if N.A. need to be explained.
			\\
			\hline
			$1$	   			& $\frac{\sqrt{2}}{b_0}\left[1 +\delta_1 \sqrt{1+\frac{2\,T}{b_0^2}} \right]^{-1/2}$ & N.A. \\
			\hline
			$-1$   			& $\left[{\frac{1}{b_0^2}-\frac{T}{2}}\right]^{-1/2}$ & N.A. \\
			\hline
			$-\frac{3}{2}$	& $\left[{\frac{\sqrt[3]{x_{-3/2}(T)}}{6^{2/3}b_0^2}-\frac{b_0^2\,T}{6^{1/3}\,\sqrt[3]{x_{-3/2}(T)}}}\right]^{-1}$ & $\sqrt{6}b_0^4\sqrt{b_0^4\,T^3+54}+18b_0^4$ \\
			\hline
			$2$	  			& $\Bigg[\frac{2^{2/3}\,b_0^{\frac{4}{3}}}{3^{1/3}\sqrt[3]{x_2(T)}}+\frac{b_0^{\frac{2}{3}}}{6^{2/3}}\,\sqrt[3]{x_2(T)}\Bigg]^{-1/2}$ & $\sqrt{3}\sqrt{27\,T^2-16b_0^2}+9\,T$ \\
			\hline
			$-2$  			& $\frac{1}{2} \sqrt{b_0^2\,T+\delta_1\,b_0\sqrt{b_0^2\,T^2+16}}$ & N.A. \\
			\hline
			$3$	  			& ${\tiny \Bigg[\frac{\delta_1}{2}\sqrt{f_1(T)}+\frac{\delta_2}{2}\Bigg[\frac{2\delta_1\,b_0^2}{\sqrt{f_1(T)}} -f_1(T)\Bigg]^{1/2}\Bigg]^{-1/2}}$ & $\sqrt{3}\sqrt{32T^3+27b_0^2}+9b_0$ \\
			& $f_1(T)=b_0\sqrt[3]{\frac{x_3(T)}{18}} -2b_0\sqrt[3]{\frac{2}{3x_3(T)}}\,T$ & \\
			\hline	
		\end{tabular}
		
	\end{table}
	%\vspace{-8pt}
	\begin{table}[h]%
		\caption{{Values} of $ G_{-1,b}\left(r(T)\right)$ special function as defined by Equation \eqref{3158function} where $a=-1$.}
		\label{tableb2}
		\begin{tabular}{cc}
			\hline
			\boldmath{$b$	}& \boldmath{$G_{-1,b}\left(r(T)\right)$}  \\
			\hline
			$\frac{1}{2}$	& $\exp\left[\frac{4\delta}{b_0}\,\left[r(T)\right]^{-\frac{1}{2}}-2\delta\,b_0\,\left[r(T)\right]^{\frac{1}{2}}\right]\Bigg[\frac{7}{2}\,b_0\,r^{-1}(T)+2\delta\,r^{-\frac{3}{2}}(T)+\delta\,b_0^2\,\,r^{-\frac{1}{2}}(T) +b_0^3\Bigg]$  \\
			\hline
			$-\frac{1}{2}$	& $\exp\left[-\frac{4\delta}{b_0}\,\left[r(T)\right]^{\frac{1}{2}}+2\delta\,b_0\,\left[r(T)\right]^{-\frac{1}{2}}\right]\Bigg[\frac{5}{2}\,b_0\,r(T)+2\delta\,r^{\frac{3}{2}}(T) +\delta\,b_0^2\,\,r^{\frac{1}{2}}(T) +b_0^3\Bigg]$   \\
			\hline
			$1$	   			& $\exp\left[\frac{2\delta}{b_0\,r(T)}-\delta\,b_0\,r(T)\right]\Bigg[\frac{4\,b_0}{r^{2}(T)}+\frac{2\delta}{r^{3}(T)} +\frac{\delta\,b_0^2}{r(T)} +b_0^3\Bigg]$   \\
			\hline
			$-1$   			& $\exp\left[\frac{\delta\,b_0}{r(T)}-\frac{2\delta}{b_0}\,r(T)\right]\Bigg[2 b_0\,r^{2}(T)+2\delta\,r^{3}(T) +\delta\,b_0^2\,r(T) +b_0^3\Bigg]$   \\
			\hline
			$-\frac{3}{2}$	& $\exp\left[\frac{2\delta\,b_0}{3\left[r(T)\right]^{\frac{3}{2}}}-\frac{4\delta}{3b_0}\left[r(T)\right]^{\frac{3}{2}}\right]\Bigg[\frac{3}{2}\,b_0\,r^{3}(T)+2\delta\,r^{\frac{9}{2}}(T) +\delta\,b_0^2\,\,r^{\frac{3}{2}}(T) +b_0^3\Bigg]$   \\
			\hline
			$2$	  			& $\exp\left[\frac{\delta}{b_0\left[r(T)\right]^{2}}-\frac{\delta\,b_0}{2}\left[r(T)\right]^{2}\right]\Bigg[\frac{5\,b_0}{r^{4}(T)}+\frac{2\delta}{r^{6}(T)} +\frac{\delta\,b_0^2}{r^{2}(T)} +b_0^3\Bigg]$   \\
			\hline
			$-2$  			& $\exp\left[\frac{\delta\,b_0}{2\left[r(T)\right]^{2}}-\frac{\delta}{b_0}\left[r(T)\right]^{2}\right]\Bigg[b_0\,r^{4}(T)+2\delta\,r^{6}(T) +\delta\,b_0^2\,r^{2}(T) +b_0^3\Bigg]$   \\
			\hline
			$3$	  			& $\exp\left[\frac{2\delta}{3 b_0 \left[r(T)\right]^{3}}-\frac{\delta\,b_0}{3}\left[r(T)\right]^{3}\right]\Bigg[\frac{6\,b_0}{r^{6}(T)}+\frac{2\delta}{r^{9}(T)} +\frac{\delta\,b_0^2}{r^{3}(T)} +b_0^3\Bigg]$   \\
			\hline		
		\end{tabular}
		
	\end{table}

	\begin{table}[h]%
		\caption{{The} Equation \eqref{3100} $r(T)$ solutions where $a=-\frac{1}{2}$. Note that $\left(\delta_1,\,\delta_2,\,\delta_3\right)=\left(\pm 1,\,\pm 1,\,\pm 1\right)$ for any positive $r(T)$.}
		\label{tableb3}
		\begin{tabular}{ccc}
			\hline
			\boldmath{$b$}	& \boldmath{$r(T)$} & \boldmath{$y_b(T)$}\\
			\hline
			$1$	   			& $3\Bigg[-\delta\,b_0+\frac{2^{2/3}\,b_0^2}{\sqrt[3]{y_1(T)}}+\frac{1}{2^{2/3}}\,\sqrt[3]{y_1(T)}\Bigg]^{-1}$ & {$-4\delta\,b_0^3+3\sqrt{3}\sqrt{27b_0^2\,T^2+8b_0^4\,T}-27\delta\,b_0\,T$} \\
			\hline
			$-1$   			& $\left[-\frac{\delta}{2b_0}\pm\sqrt{\frac{1}{4b_0^2}-\frac{T}{2}}\right]^{-1}$ & N.A. \\
			\hline
			$2$	  			& $\left[{-\frac{\delta\,b_0}{2}+\delta_1\,\sqrt{\frac{b_0^2}{4}-\frac{\delta\,b_0\,T}{2}}}\right]^{-1/2}$ & N.A. \\
			\hline
			$-2$  			& $\left[-\frac{{2b_0}}{{b_0\,T+2\delta}}\right]^{1/2}$ & N.A. \\
			\hline
			$-3$	  		& $\frac{1}{6}\Bigg[\sqrt[3]{y_{-3}(T)}+\frac{b_0^2\,T^2}{\sqrt[3]{y_{-3}(T)}}-\delta\,b_0\,T\Bigg]$ & $-108\delta\,b_0-\delta\,b_0^3\,T^3+6^{3/2}\,b_0\sqrt{54+b_0^2\,T^3}$ \\
			\hline
			$4$	   			& $\Bigg[\frac{b_0^{\frac{1}{3}}}{6^{\frac{2}{3}}}\,\sqrt[3]{y_4(T)}	-\frac{\delta\,b_0^{\frac{2}{3}}\,2^{\frac{2}{3}}}{3^{\frac{1}{3}}\sqrt[3]{y_4(T)} }\Bigg]^{-1/2}$ & $\sqrt{3}\sqrt{16\delta b_0+27\,T^2}-9\delta\,T$ \\
			\hline
			$-4$   			& $\sqrt{-\frac{\delta\,b_0\,T}{4}+\delta_1\sqrt{\frac{b_0^2\,T^2}{16}-\delta\,b_0}}$ & N.A. \\
			\hline
			$6$	  			& ${\tiny \frac{1}{\sqrt{\frac{\delta_1}{2}\sqrt{f_2(T)} +\frac{\delta_2}{2} \sqrt{-f_2(T)-\frac{\delta\,\delta_2}{2b_0\sqrt{f_2(T)}}}}}}$ & $9\delta\,b_0^2+\sqrt{3}\,b_0\sqrt{27b_0^2-32\delta\,b_0\,T^3}$ \\
			& $f_2(T)=2\sqrt[3]{\frac{2}{3y_6(T)}}b_0\,T +\frac{\delta}{\sqrt[3]{18}}\left(y_6(T)\right)^{1/3}$ &  \\
			\hline
			$-6$  			& $\left[\frac{\sqrt[3]{y_{-6}(T)}}{6^{2/3}}-\frac{\delta\,b_0\,T}{\sqrt[3]{6y_{-6}(T)}}\right]^{1/2}$ & $\sqrt{6}\,b_0\sqrt{\delta\,b_0\,T^3+54}-18\delta\,b_0$ \\
			\hline
			$-8$	  		& $\sqrt{\frac{\delta_1}{2}\sqrt{f_3(T)}+\frac{\delta_2}{2}\sqrt{-f_3(T)-\frac{\delta_1\delta\,b_0\,T}{\sqrt{f_3(T)}}}}$ & $\sqrt{3}b_0\sqrt{27b_0^2\,T^4-4096\delta\,b_0}+9b_0^2\,T^2$ \\
			& $f_3(T)=\frac{8\delta\,b_0}{\sqrt[3]{3y_{-8}(T)}}+\frac{1}{2\cdot\,3^{2/3}}\left(y_{-8}(T)\right)^{1/3}$ & \\		
			\hline		
		\end{tabular}
		
	\end{table}

	\begin{table}[h]%
		\caption{{Values} of $ G_{-\frac{1}{2},b}\left(r(T)\right)$ special function as defined by Equation \eqref{3247function} where $a=-\frac{1}{2}$.}
		\label{tableb4}
		\begin{tabular}{cc}
			\hline
			\boldmath{$b$}	& \boldmath{$G_{-\frac{1}{2},b}\left(r(T)\right)$} \\
			\hline
			$1$	   			&  $-\frac{\delta\,b_0}{2}\sqrt{r(T)}\left(r^{-1}(T) -2\delta\,b_0\right)\exp\left[-\delta\,b_0\,r(T)\right]$  \\
			\hline
			$-1$   			& $\frac{-\exp\left[\frac{\delta\,b_0}{r(T)}\right]}{\left(r(T)+2\delta\,b_0\right)^2}\sqrt{r(T)}\Bigg[\frac{3}{2}r^2(T)+{\delta\,b_0}\,r(T) -\frac{2\left[\frac{3}{4}\,r^{2}(T)+b_0^2\right]}{\left[r(T)+2\delta\,b_0\right]} \left(r(T)+\delta\,b_0\right)\Bigg] $   \\
			\hline
			$2$	  			& $\frac{\exp\left[-\frac{\delta\,b_0}{2} r^2(T)\right]}{2\left(1+2\delta\,b_0\,r^2(T)\right)^{\frac{1}{2}}} \Bigg[\frac{\left(3-2\delta\,b_0r^2(T)\right)}{r^{5/2}(T)}-\frac{\left[3r^{-4}(T)-4b_0^2\right]}{\left[r^{-2}(T)+2\delta\,b_0\right]}{\left(r^{-2}(T)+\delta\,b_0\right)}\left[r(T)\right]^{3/2}\Bigg]$   \\
			\hline
			$-2$  			&  $\frac{-\left[r(T)\right]^{1/2}\exp\left[\frac{\delta\,b_0}{2r^2(T)}\right]}{2\left(r^2(T)+2\delta\,b_0\right)^{3/2}}\Bigg[5r^{4}(T)+2\delta\,b_0\,r^2(T)
			-\frac{\left[5r^{4}(T)+4b_0^2\right]}{\left[r^{2}(T)+2\delta\,b_0\right]}\left(r^{2}(T)+\delta\,b_0\right)\Bigg]$  \\
			\hline
			$-3$	  		&  $\frac{-r^{\frac{7}{2}}(T)\,\exp\left[\frac{\delta\,b_0}{3r^3(T)}\right]}{2\left(r^3(T)+2\delta\,b_0\right)^{\frac{4}{3}}}\Bigg[\left(7r^{3}(T)+2\delta\,b_0\right)  -\frac{\left[7r^{6}(T)+4b_0^2\right]}{\left[r^{3}(T)+2\delta\,b_0\right]}\frac{\left(r^{3}(T)+\delta\,b_0\right)}{r^{3}(T)}\Bigg]$  \\
			\hline
			$4$	   			&  $\frac{\exp\left[-\frac{\delta\,b_0}{4}\,r^4(T)\right]}{2\left[r(T)\right]^{\frac{9}{2}}\left(1+2\delta\,b_0 r^4(T)\right)^{\frac{3}{4}}}\Bigg[\left[7-\delta\,b_0 r^4(T)\right]  -\frac{\left[7-4b_0^2r^8(T)\right]}{\left[1+2\delta\,b_0 r^4(T)\right]}\left[1+\delta\,b_0 r^4(T)\right]\Bigg]$  \\
			\hline
			$-4$   			&  $\frac{-\left[r(T)\right]^{\frac{9}{2}}\exp\left[\frac{\delta\,b_0}{4r^4(T)}\right]}{\left(r^4(T)+2\delta\,b_0\right)^{\frac{5}{4}}}\Bigg[\frac{9}{2}r^{4}(T)+\delta\,b_0 -\frac{\left[\frac{9}{4}\,r^{8}(T)+b_0^2\right]}{\left[\frac{1}{2}\,r^{4}(T)+\delta\,b_0\right]}\frac{\left(r^{4}(T)+\delta\,b_0\right)}{r^4(T)}\Bigg]$  \\
			\hline
			$6$	  			&  $\frac{\exp\left[-\frac{\delta\,b_0}{6}r^6(T)\right]}{2\left[r(T)\right]^{\frac{13}{2}}\left(1+2\delta\,b_0 r^6(T)\right)^{\frac{5}{6}}}
			\Bigg[11-2\delta\,b_0 r^{6}(T) -\frac{\left[11-4b_0^2 r^{12}(T)\right]}{\left[1+2\delta\,b_0 r^6(T)\right]}\left(1+\delta\,b_0 r^6(T)\right)\Bigg]$  \\
			\hline
			$-6$  			&  $\frac{-\left[r(T)\right]^{\frac{13}{2}}\exp\left[\frac{\delta\,b_0}{6r^6(T)}\right]}{2\left(r^6(T)+2\delta\,b_0\right)^{\frac{7}{6}}}
			\Bigg[13 r^{6}(T)+2\delta\,b_0 -\frac{\left[13 r^{12}(T)+4b_0^2\right]}{\left[r^{6}(T)+2\delta\,b_0\right]}\frac{\left(r^{6}(T)+\delta\,b_0\right)}{r^{6}(T)}\Bigg]$  \\
			\hline
			$-8$	  		& $\frac{-\left[r(T)\right]^{\frac{17}{2}}\exp\left[\frac{\delta\,b_0}{8 r^8(T)}\right]}{2\left(r^8(T)+2\delta\,b_0\right)^{\frac{9}{8}}}\Bigg[17 r^{8}(T)+2\delta\,b_0 -\frac{\left[17 r^{16}(T)+4b_0^2\right]}{\left[r^{8}(T)+2\delta\,b_0\right]}\frac{\left(r^{8}(T)+\delta\,b_0\right)}{r^{8}(T)}\Bigg]$   \\		
			\hline			
		\end{tabular}
		
	\end{table}
	
}

\end{document}